\documentclass[USenglish]{article}
\pdfoutput=1

\usepackage[utf8]{inputenc}
\usepackage[small]{dgruyter}
\usepackage{microtype}
\usepackage{natbib}
\usepackage{xcolor}

\def \simulator { f }
\def \simulatorx { f(x) }

\def \expectation { \textrm{E} }
\def \variance { \textrm{Var} }
\def \covariance { \textrm{Cov} }

\def \dist { \rho }

\def \obs { z }
\def \md { \epsilon }
\def \me { e }

\def \independent {\perp\!\!\!\perp}

\def \bestx { x^\star }

\def \imp { I }
\def \setx { \mathcal{X} }
\def \bestset { \setx^\star }

\def \rv { W }

\def \det {\textrm {det}}

\newcommand{\rev}[1]{\textcolor{black}{#1}}
\newcommand{\ian}[1]{\textcolor{black}{#1}}

\usepackage[refpage]{nomencl}

\makenomenclature

\begin{document}

  \articletype{Original Research Article}

  \author*[1]{Samuel E. Jackson}
  \author[2]{Ian Vernon}
  \author[3]{Junli Liu} 
  \author[4]{Keith Lindsey}
  \affil[1]{Southampton Statistical Sciences Research Institute, University of Southampton, Southampton, UK, email: s.e.jackson@soton.ac.uk}
  \affil[2]{Department of Mathematical Sciences, Durham University, Durham, UK}
  \affil[3]{School of Biological and Biomedical Sciences, Durham University, Durham, UK}
  \affil[4]{School of Biological and Biomedical Sciences, Durham University, Durham, UK}
  \title{\rev{Understanding hormonal crosstalk in Arabidopsis root development via  \ian{Emulation and} History Matching}}
  \runningtitle{Links between model parameter spaces and measurements}
  \abstract{A major challenge in plant developmental biology is to understand how plant growth is coordinated by interacting hormones and genes. To meet this challenge, it is important to not only use experimental data, but also formulate a mathematical model. For the mathematical model to best describe the true biological system, it is necessary to understand the parameter space of the model, along with the links between the model, the parameter space and experimental observations. We develop sequential history matching methodology, using Bayesian emulation, to gain 
\rev{substantial} insight into biological model parameter spaces. This is achieved by finding sets of acceptable parameters in accordance with successive sets of physical observations. These methods are then applied to a complex hormonal crosstalk model for Arabidopsis root growth. In this application, we demonstrate how an initial set of 22 observed trends reduce \rev{the volume of} the set of acceptable inputs to \rev{a proportion of $ 6.1 \times 10^{-7} $} of the original space. Additional sets of biologically relevant experimental data, \rev{each of size 5,} reduce the size of this space by a further three and two orders of magnitude respectively. Hence, we provide insight into the constraints placed upon the model structure by, and the biological consequences of, measuring subsets of observations.
}
  \journalyear{...}
  \journalvolume{..}
  \journalissue{..}
  \startpage{1}
  \aop

\maketitle

\section{Background}


\subsection{Use and Understanding of Scientific Models in Systems Biology}

One of the major challenges in biology is to understand how functions in cells emerge from molecular components. Computational and mathematical modelling is a key element in systems biology which enables the analysis of biological functions resulting from non-linear interactions of molecular components.  The kinetics of each biological reaction can be systematically represented using a set of differential equations \citep{SBPF, MEAHCA, TKMBN, FGSKMPGA, TGSKCM}.  Due to the multitude of cell components and the complexity of molecular interactions, the kinetic models often involve large numbers of reaction rate parameters, that is parameters representing the rates at which reactions encapsulated by the model are occurring \citep{PMMEP, SMHCA, MPHG}.  Quantitative experimental measurements can be used to formulate the kinetic equations and learn about the associated rate parameters \citep{SBPF, MEAHCA, PMMEP, MMPB, IPPHCARD}.  This in turn provides insight into the functions of the actual biological system.

An important question is therefore how much information about the kinetic equations and parameters can be obtained from an experimental measurement.  Since a key aspect of experimental measurements in modern biological science is the study of the functions of specific genes, the answer to the above question is also important for understanding the role of each gene within the components of a biological system.

In plant developmental biology, a major challenge is to understand how plant growth is coordinated by interacting hormones and genes. Previously, a hormonal crosstalk network model - which describes how three hormones (auxin, ethylene and cytokinin) and the associated genes coordinate to regulate Arabidopsis root development - was constructed by iteratively combining regulatory relationships derived from experimental data and mathematical modelling \citep{MEAHCA, IPPHCARD, SMHCA, MPHG, MAPCC, RPPI}. 
However, for the mathematical model to best link with Arabidopsis root development, it is necessary to understand the parameter space of the model and identify all acceptable parameter combinations.  Little is known about how acceptable parameter \rev{combinations} of a model \rev{can be identified in light of} specific experimental data.  Therefore, this work explores \rev{how} the acceptable parameter space of a complex model of hormonal crosstalk \citep{IPPHCARD, SMHCA, MPHG, MAPCC, RPPI} \rev{is assessed given} experimental measurements by employing Bayesian history matching techniques \rev{\citep{CSBS, GFBUA, BLUAORBMCE, AHMMR, SSHMSS}}.  
Additionally, we utilise these techniques to analyse how learning about the functions of a gene through particular relevant experiments can inform us about acceptable model parameter space.

\subsection{Efficient Analysis of Scientific Models}

Complex systems biology models are frequently high dimensional and can take a substantial amount of time to evaluate \citep{SMHCA}, thus comprehensive analysis of the entire input space, requiring vast numbers of model evaluations, may be unfeasible \citep{BUCSBM}.  We are frequently interested, as is the case in this paper, in comparing the scientific model to observed data (usually measured with uncertainty), necessitating a possibly high dimensional parameter search.  Our history matching approach aims to find the set of all possible combinations of input rate parameters which could have plausibly led to the observed data, given all sources of uncertainty involved with the model and the experimental data \citep{CSBS, GFBUA,BUCSBM}.  This biologically relevant aim requires comprehensive exploration of the model's behaviour over the whole input space, and therefore efficient techniques, such as emulation \rev{\citep{CSBS, BCCM, GFDEMEP, GPML, HMERCMPS}}, are required.  An emulator mimics the biological model, but is substantially faster to evaluate, hence facilitating the large numbers of evaluations that are needed.

We are often keen to understand the contribution of particular sets of observations towards being able to answer critical scientific questions.  Sequential incorporation of datasets into a history matching procedure, as presented in this article, is very natural and can allow us to attain such understanding.  Comprehensive understanding and parameter searching of the hormonal crosstalk model for Arabidopsis root development \citep{IPPHCARD}, by sequentially history matching specific groups of experimental observations, is the focus of this paper.

\section{Methods}


\subsection{Bayes Linear Emulation \label{BLE}}

In this section we \rev{review} the process of constructing an emulator for a complex systems biology model.  For more detail see \citet{BUCSBM}.  We represent the set of input rate parameters of the model as a vector $ x $ of length $ d $, and the outputs of the model as vector $ f(x) $ of length $ q $.

\nomenclature{$ f $}{simulator}
\nomenclature{$ x $}{simulator input vector}
\nomenclature{$ q $}{dimension of simulator input vector}
\nomenclature{$ d $}{dimension of simulator output vector}

A Bayes linear emulator is a fast statistical approximation of the systems biology model built using a set of model runs, providing an expected value for the model output at a particular point $ x $, along with a corresponding uncertainty estimate reflecting our beliefs about the uncertainty in the approximation \citep{BLA, BLS}.  The main advantage of emulation is its computational efficiency: often an emulator is several orders of magnitude faster to evaluate than the model it is mimicking. Emulation has been successfully applied across a variety of scientific disciplines such as climate science \citep{SECMP, GFDEMEP}, cosmology \citep{PSGF, CUII}, epidemiology \citep{BECDEMI}, \rev{humanitarian relief \citep{MECSMSD}}, as well as systems biology \citep{BUCSBM}.

\rev{We index by $ i = 1,...,q $ the output components of the model.  Each output component of the model $ f_i(x) $ can be represented} in emulator form as presented by \citet{GFBUA}: \rev{
\begin{equation}
f_i(x) = \sum_{j=1}^{J_i} \beta_{ij} g_{ij}(x_{A_i}) + u_i(x_{A_i}) + w_i(x) \label{emulator2}
\end{equation}}
where $ x_{A_i} $ represents the subset of active variables, that is the input \rev{components of} $ x $ which are most influential for output $ f_i(x) $, $ g_{ij} $ are \rev{$J_i$} known functions of $ x_{A_i} $, and $ \beta_{ij} $ are the corresponding coefficients to the regression functions $ g_{ij} $.  $ u_i(x_{A_i}) $ \rev{is a second-order weakly stationary stochastic process which} captures residual variation in $ x_{A_i} $.  \rev{\emph{A priori}, we assume that $ \expectation[u_i(x_{A_i})] = 0 $, along with the following} covariance structure: 
\begin{equation}
\covariance[u_i(x_{A_i}), u_i(x^\prime_{A_i})]  =  \sigma_{u_i}^2 \exp \left( - \sum_{k\in S_{A_i}}  \left\{ \frac{x_k - x_k^\prime}{\theta_{ik}} \right\}^2 \right) \label{Cov}
\end{equation}
where $S_{A_i}$ is the set of indices of the active inputs for output $i$. 
$ w_i(x) $ is a zero-mean ``nugget'', \rev{or residual error}, term with constant variance $ \sigma_{w_i}^2 $ over $ x $ and $ \covariance[w_i(x), w_i(x^\prime)] = 0 $ for $ x \neq x^\prime $.  The nugget represents the effects of the remaining inactive input variables \citep{GFBUA}.  \rev{We also make the assumption that: $$ \covariance[\beta_{ij}, u_i(x_{A_i})] = \covariance[\beta_{ij}, w_i(x)] = \covariance[u_i(x_{A_i}), w_i(x)] = 0 $$} for all $ i,j$.    \rev{The parameters in Equation (\ref{Cov}) should, in principle, be specified \emph{a priori}, however, various techniques are available for estimating them from the data.  For example, we can use maximum likelihood estimates if we are happy to specify distributions \citep{PEGPE}, we can use simple heuristics \citep{GFBUA}, or we can use predictive diagnostics, such as Leave-One-Out-Cross-Validation \citep{CVEHPGP}.  However they are chosen, the resulting choices should be checked using rigorous diagnostics.}

\nomenclature{$ x_{A_i} $}{set of active variables}
\nomenclature{$ g_{ij} $}{known functions of $ x_{A_i} $}
\nomenclature{$ u_i(x_{A_i}) $}{second-order weakly stationary process}
\nomenclature{$ w_i(x) $}{zero-mean ``nugget'' term}
\nomenclature{$ S_{A_i} $}{set of indices of the active inputs for output $ i $}
\nomenclature{$ \sigma_{w_i}^2 $}{constant variance of nugget term}

Suppose $ D_i = (f_i(x^{(1)}),..., f_i(x^{(n)})) $ represents model output component $ i $ evaluated at $n$ model runs performed at locations $ x^{(1)},....,x^{(n)}  $.  The Bayes linear emulator output for simulator output component $ i $ at a new $ x $ is given by \ian{the Bayes linear update formulae} \citep{BLA, BLS}:
\begin{eqnarray}
\expectation_{D_i}[f_i(x)] & =   & \expectation[f_i(x)] + \covariance[f_i(x), D_i] \variance[D_i]^{-1} (D_i - \expectation[D_i])  \label{Exp} \\
\variance_{D_i}[f_i(x)] & =  & \variance[f_i(x)] - \covariance[f_i(x), D_i] \variance[D_i]^{-1} \covariance[D_i, f_i(x)] \label{Var}
\end{eqnarray}
where the notation $ \expectation_{D_i}[f_i(x)] $ and $ \variance_{D_i}[f_i(x)] $ reflects the fact that we have adjusted our prior beliefs about $ f_i(x) $ by model runs $ D_i $, and can be obtained for any point $ x $ using Equations (\ref{emulator2}) and (\ref{Cov}).  \rev{We note that in the literature it is common to assume normal and Gaussian process priors for $ \beta $ and $ u(x) $ in Equation (\ref{emulator2}) \citep{BCCM, GPEDCC, GPESOMCS}, thus resulting in Gaussian process emulation.   In this case, the resulting Bayesian update equations are practically similar to Equations (\ref{Exp}) and (\ref{Var}) presented above, however, methodologically involve additional distributional assumptions \ian{that may be harder to justify, thus requiring stricter corresponding diagnostics to be satisfied, and affect the resulting inference.  We would rather go as far as possible without making such distributional assumptions.  In this spirit, the Bayes linear framework is more similar to traditional kriging \citep{MG, GPML}, noting that this term is now sometimes used to mean several different related approaches.  Having said that, we note that kriging is derived from classical unbiased estimator arguments, whereas the Bayes linear paradigm follows from a foundational position, following DeFinetti \citep{TP, TP2}, that treats expectation as primitive and does not invoke concepts such as unbiasedness. The Bayes linear paradigm has been applied in a wide range of scenarios \citep{SPDF, BLERRM, BLAWERA}: for example, in the context of the emulation of computer models, it has allowed tractable multilevel emulation due to multi-fidelity models, thus going beyond standard universal kriging \citep{CSBS, MLE}. }}

\nomenclature{$ D_i $}{set of simulator runs}
\nomenclature{$ U_i(x) $}{standardised prediction errors}

Emulator design is the process of selecting the points in the input space $ x^{(1)}, ..., x^{(n)} $ at which the simulator will be run in order to construct an emulator \citep{DACE}.  A popular design choice in the computer model literature is the Maximin Latin Hypercube design \citep{CTM, BPDF}, \rev{however, other options are also available (see, for example, \citet{DOE, DAE})}.  

Performing emulator diagnostics \citep{DGPE}, for example calculating standardised prediction errors: 
\begin{equation} 
U_i(x) = \frac{f_i(x) - \expectation_{D_i}[f_i(x)]}{\sqrt{\variance_{D_i}[f_i(x)]}} \label{EmDiag} 
\end{equation} 
for a set of validation data, is essential for validating an emulator.  Large errors $ U_i(x) $ indicate conflict between simulator and emulator.  \rev{If these are observed, the emulator is not valid for inference.  It may be possible that the emulator prior beliefs were misspecified, for example, as a result of incorrect prior specifications for the parameters $ \beta $, $ \sigma_u^2 $ and $ \theta $.  Alternatively, it could be indication of an erratically behaved model that would require substantially more model runs in order to be emulated well.}

\subsubsection*{1 Dimensional Example}

\begin{figure*}
\includegraphics[width=12cm]{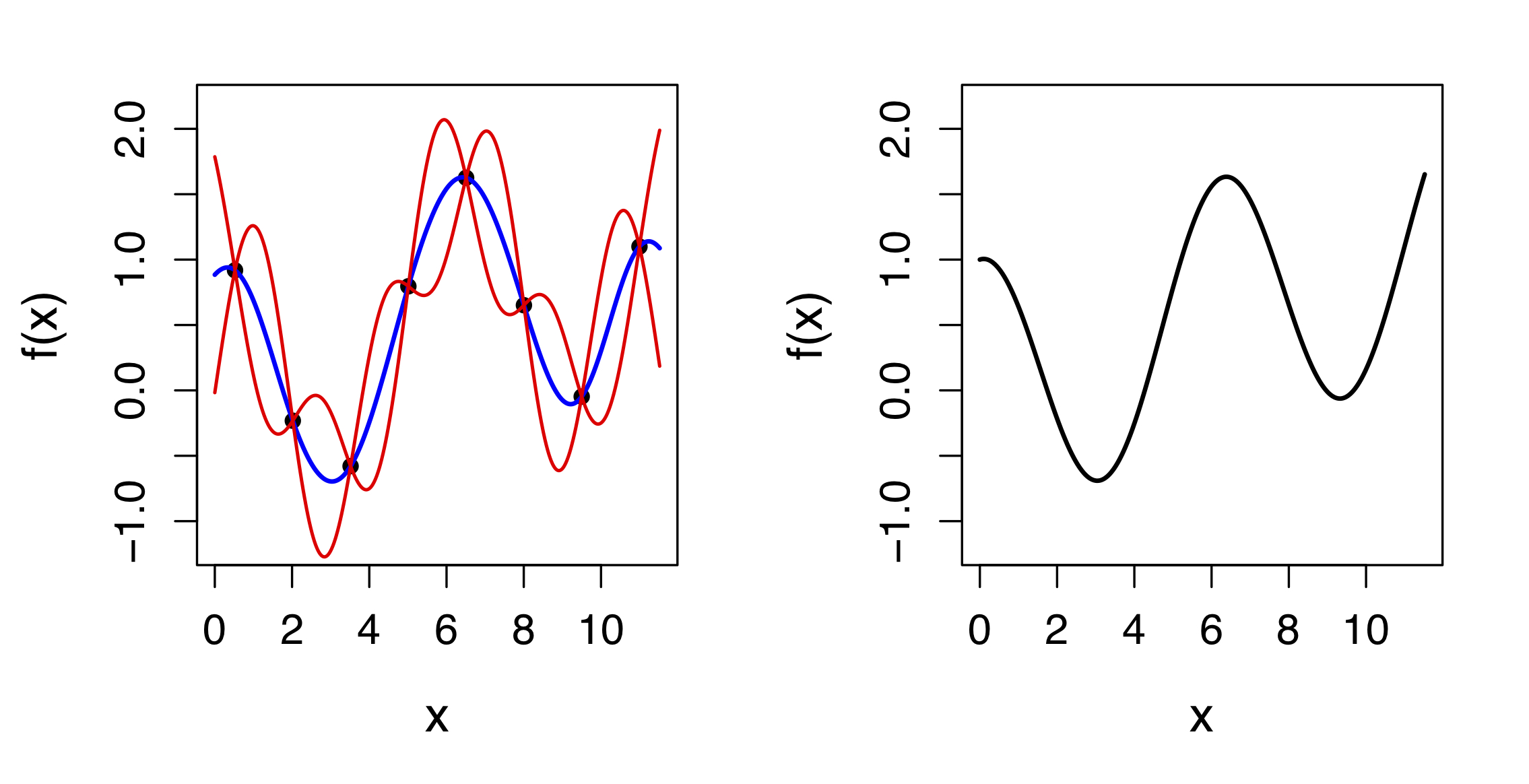}
\caption[1D Emulator Example]{Left: Emulator expectation $\expectation_{D}[f(x)]$ (blue) with emulator credible intervals $ \expectation_{D}[f(x)] \pm 
3\sqrt{\variance_{D}[f(x)] } $ (red) for an emulator of $ f(x) = 0.1x + \cos(x) $ constructed using 8 training points.  Right: Simulator function $ f(x) = 0.1x + \cos(x) $.  \label{1DEmEx} }
\end{figure*}

In this section we demonstrate emulation techniques on a simple one-dimensional example.  We will suppose that we wish to emulate the simple function $ \simulatorx = 0.1x + \cos(x) $ in the range $ [0, \frac{11\pi}{3}] $, where we treat $ x $ as a rate parameter that we wish to learn about, and $ \simulatorx $ as a chemical concentration that we could measure.  We assume an emulator of the form given by Equation (\ref{emulator2}) with covariance structure given by Equation (\ref{Cov}).  We assume a zero mean function, \rev{that is, $ g^T(x)\beta = 0 $}, and that $ \sigma^2_u = 0.5 $, $ \theta =1.5 $ and $ \sigma^2_w = 0 $.  
We also specify a prior expectation $ \expectation[f(x)] = 0 $.  Having specified our prior beliefs, we then use the update rules given by Equations (\ref{Exp}) 
and (\ref{Var}) to obtain the adjusted expectation $\expectation_{D}[f(x)]$ and variance $\variance_{D}[f(x)]$ for $ f(x) $.  The results of this emulation 
process are shown in the left panel of Figure \ref{1DEmEx}.  The blue lines represent the emulator expectation $ \expectation_{D}[f(x)]$ of the simulator 
output for the test points.  The red lines represent the emulator mean $\pm 3$ emulator standard deviations, given as $ \expectation_{D}[f(x)] \pm 
3\sqrt{\variance_{D}[f(x)] } $, \rev{these being bounds for a 95\% credible interval, following Pukelsheim's $ 3 \sigma $ rule \citep{3SR}, \ian{which states that at least 95\% of the probability mass of any unimodal continuous distribution will lie within $ \pm 3 $ standard deviations of the mean, regardless of asymmetry or skew}}.  
By comparison with the right panel of Figure \ref{1DEmEx}, we can see that the emulator estimates the simulator output well, with some uncertainty.  We note that we would not expect such large emulator uncertainty on such a smooth function as this, but have deliberately ensured that there is a large uncertainty for illustrative purposes, and in particular to highlight the effects of additional runs on reducing emulator uncertainty in the continuation of this example in Section \ref{HM}.

\subsection{History Matching \label{HM}}

History matching concerns the problem of finding the set of inputs to a model for which the corresponding model outputs give acceptable matches to observed historical data, given our state of uncertainty about the model itself and the measurements.  History matching has been successfully applied across many scientific disciplines including oil reservoir modelling \rev{\citep{BLSMHRH, CSBS, SSBD, BLUAORBMCE, RPHM}}, \rev{engineering \citep{SSHMSS, BHMFMDSHM}}, epidemiology \citep{BHMCIDM, HMHIV, Yiannis_HIV_2, McCreesh2017}, climate modelling \citep{HMERCMPS} and systems biology \citep{BUCSBM}. Here we provide a brief summary of the history matching procedure (see \citet{GFBUA, BUCSBM} for more details). 

We need a general structure to describe the link between a complex model and the corresponding physical system.  We use the direct simulator approach, otherwise known as the best input approach \citep{BLCPCS}, where we \rev{posit} that there exists a value $ x^\star $ such that $ \simulator(\bestx) $ best represents the real biological system, which we denote as $y$ \citep{BLCPCS, RBM}. We then formally link the $i$th output of the model to the $i$th real system value $y_i$ via
\begin{equation} 
y_i = \simulator_i(\bestx) + \md_i \label{breakdown1}
\end{equation}
and link the experimental observation $\obs_i$ corresponding to output $ i $ to the real system via
\begin{equation} 
z_i = y_i + \me_i \label{breakdown2}
\end{equation}
where we assume $ \simulator_i(\bestx) \independent \md_i \independent \me_i $, with $ a \independent b $ indicating that random variables $ a $ and $ b $ are uncorrelated \citep{BLS}.
Here, $ \simulator_i(\bestx) $ is the model run at best input $ \bestx $, $ \md_i $ is a random variable which reflects our uncertainty due to discrepancy between the model run at the best possible input combination setting and the real world \rev{\citep{BCCM, LAPP, AMA, QMUCMDI}}, and $ \me _i $ is a random variable which incorporates our beliefs about the error between each desired real world quantity and its corresponding observed measurement.  We assume $ \expectation[\md_i] = \expectation[\me_i] = 0 $, $ \variance[\md_i] = \sigma_{\md_i}^2 $ and $ \variance[\me_i] = \sigma_{\me_i}^2 $.  The connection between system, observation and model given by (\ref{breakdown1}) and (\ref{breakdown2}) is simple but well-used \citep{CSBS, BLCPCS, BHMCIDM}, and judged sufficient for our purposes.  For discussion of more advanced approaches see \citet{RBM}.

\nomenclature{$ y_i $}{physical system value}
\nomenclature{$ x^\star $}{best input}
\nomenclature{$ \epsilon_i $}{model discrepancy}
\nomenclature{$ e_i $}{measurement error}
\nomenclature{$ z_i $}{observed value}

We then aim to find the set $ \bestset $ of all input combinations $x$ that are consistent with Equations~(\ref{breakdown1}) and (\ref{breakdown2}), that is those that
will provide acceptable matches between model output and data.  \rev{Note that classifying points in this way can lead to $ \bestset $ being empty, \ian{an informative conclusion which would contradict the posited existence of $ \bestx $ in Equation (\ref{breakdown1}), and imply that the model may not be fit for purpose}}.
To analyse whether a point $ x \in \bestset $ it is practical to use implausibility measures for each output $ i $, as given, for example, in \citet{BLSMHRH, CSBS,GFBUA}: 
\begin{equation} 
\imp^2_i(x) = \frac{(\expectation_{D_i}[\simulator_i(x)] - \obs_i)^2}{\variance_{D_i}[\simulator_i(x)] + \sigma^2_{\md_i} + \sigma^2_{\me_i}} \label{Implausibility} 
\end{equation}  
If $ \imp_i(x) $ is large this suggests that we would be unlikely to obtain an acceptable match between model output and observed data were we to run the model at $x$.   This is after taking into account all the uncertainties associated with the model and the measurements.  We develop a combined implausibility measure over multiple outputs such as $ \imp_M(x) = \max_i \imp_i(x) $, $ \imp_{2M}(x) =  \max_i ( \{\imp_i(x) \} \setminus \imp_M(x) ) $ and $ \imp_{3M}(x) = \max_i ( \{\imp_i(x) \} \setminus \{ \imp_M(x), \imp_{2M}(x) \} ) $ \citep{GFBUA}.  We class $ x $ as implausible if the values of these measures lie above suitable cutoff thresholds \citep{CSBS, GFBUA}.

\nomenclature{$ \mathcal{X} $}{non-implausible set}
\nomenclature{$ I_i(x) $}{implausibility function}
\nomenclature{$ Q_k $}{set of outputs emulated at wave $ k $}
\nomenclature{$ \mathcal{X}_k $}{non-implausible set after wave $ k $ of a history match}

History matching using emulators proceeds as a series of iterations, called waves, discarding regions of the input parameter space at each wave.  At the $ k $th wave emulators are constructed for a selection of well-behaved outputs $ Q_k $ over the non-implausible space remaining after wave $ k-1 $.  These emulators are used to assess implausibility over this space where points with sufficiently large values are discarded to leave a smaller set $ \setx_k $ remaining 
\citep{GFBUA,BUCSBM}.

The history matching algorithm is as follows:
\begin{enumerate}
\item \rev{Let $ \setx_0 $ be the initial domain space of interest and set $ k = 1 $.}
\item Generate a design for a set of runs over the non-implausible space $ \setx_{k-1} $, for example using a maximin Latin hypercube with rejection \citep{GFBUA}.
\item Check to see if there are new, informative outputs that can now be emulated accurately and add them to the previous set $ Q_{k-1} $ to define $ Q_k $.
\item Use the design of runs to construct new, more accurate emulators defined only over $ \setx_{k-1} $ for each output in $ Q_k $.
\item Calculate implausibility measures over $ \setx_{k-1} $ for each of the outputs in $ Q_k $.
\item Discard points in $ \setx_{k-1} $ with $ I(x) > c $ to define a smaller non-implausible region $ \setx_k $.
\item If the current non-implausible space $ \setx_k $ is sufficiently small, go on to step \rev{8}.  Otherwise repeat the algorithm from step \rev{2} for wave $ k + 1 $.  The non-implausible space is sufficiently small if it is empty or if the emulator variances are small in comparison to the other sources of uncertainty 
($\sigma^2_{\md}$ and $\sigma^2_{\me}$), since in this case more accurate emulators would do little to reduce the non-implausible space further.  
\item Generate a large number of acceptable runs from $ \setx_k $, sampled according to scientific goal.
\end{enumerate}
It should be the case that $ \bestset \subseteq \setx_k \subseteq \setx_{k-1} $ for all $ k $, where $ \bestset = \{x : \max_i \imp_i(x) < c \} $ for some threshold $ c $, where each $ \imp_i(x) $ is calculated using expression (\ref{Implausibility}) with $ \expectation[\simulator_i(x)] = \simulator_i(x) $ and $ \variance[\simulator_i(x)] = 0 $, that is were we to know the simulator output everywhere. The choice of cutoff $c=3$ is frequently chosen, motivated by Pukelsheim's 3-sigma rule~\citep{3SR}, \rev{which in this case implies that $ P(I_i(x) < 3 | x = \bestx) > 0.95 $ for any unimodal continuous distribution for the combined error term \ian{
$ \epsilon_i + e_i +  (\simulator_i(x) -  \expectation[\simulator_i(x)])  $}}.
 This iterative procedure is powerful as it quickly discards large regions of the input space as implausible based on a small number of well behaved (and hence easy to emulate) outputs. In later waves, outputs that were initially hard to emulate, possibly due to their erratic behaviour in uninteresting parts of the input space, become easier to emulate over the much reduced space $ \setx_k $.
Careful consideration of the initial non-implausible space $ \setx_0 $ is important.  It should be large enough such that no potentially scientifically interesting input combinations are excluded.  A more in-depth discussion of the benefits of this history matching approach, especially in problems requiring the use of emulators, may be found in \citet{BUCSBM}.  \rev{In addition, a comparison of using Bayes linear emulators and Gaussian process emulators within a history match can be found in \citet{RGFBUA}}.


\subsubsection*{1 Dimensional Example Continued}

\begin{figure*}[h]
\includegraphics[width=12cm]{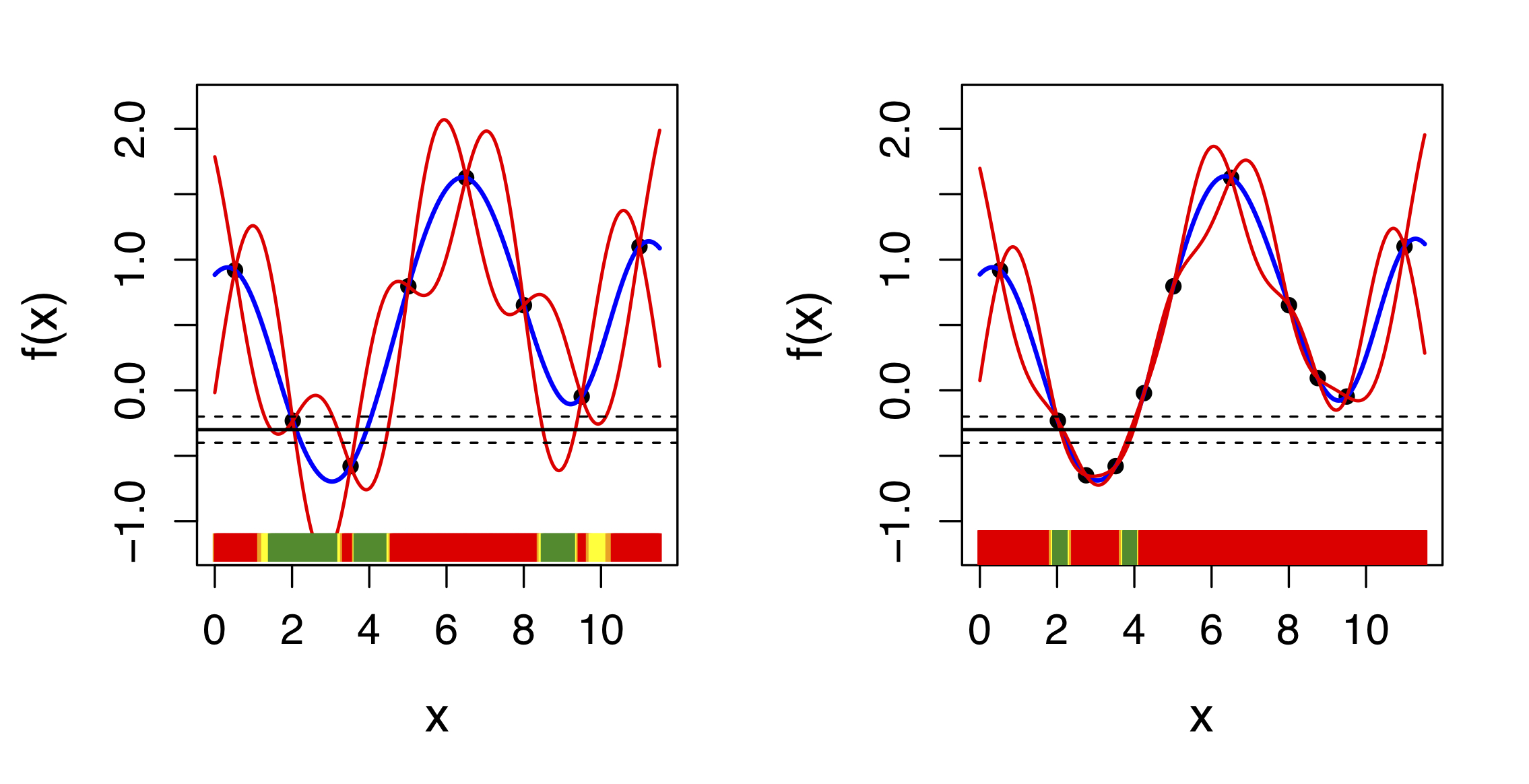}
\caption[1D History Matching Example]{Left panel: Emulators for the simple 1D example $ \simulatorx = 0.1x + \cos(x) $ as given by the left panel of Figure \ref{1DEmEx}.  The blue line represents the emulator's updated expectation $ \expectation_D[\simulatorx] $ and the pair of red lines give the credible interval $ \expectation_D[\simulatorx] \pm 3\sqrt{\variance_D[\simulatorx]} $.  Observation $ z $ along with observed error are shown as horizontal black solid and dashed lines respectively.  The implausibilities $ \imp(x) $ are represented by the colours on the x-axis, with red representing high implausibility, orange and yellow representing borderline implausibility, and green representing low implausibility ($ \imp(x) < 3 $).  Right panel: The wave 2 emulator for the same function, now including three additional wave 2 runs. \label{HMEx}}
\end{figure*}

Figure \ref{HMEx} (left panel) shows the emulator expectation and \rev{conservative bounds for the 95\% credible intervals}, as given by Figure \ref{1DEmEx} (left panel), however, now an observation $ z = -0.3 $ along with observed error is included as solid and dashed lines respectively.  In this example, we let model discrepancy be 0, and set the measurement standard error $ \sigma_e = 0.05 $.  Along the bottom of the figure is the implausibility $ \imp(x) $ for each $ x $ value represented by colour: red for large implausibility values, orange and yellow for borderline implausibility, and green for low implausibility 
($ \imp(x) < 3 $) \citep{3SR}.

$ \setx_0 $ is the full initial range of $ x $, that is $ 0 \leq x \leq \frac{11\pi}{3} $.  $ \setx_1 $ is as shown by the green regions in Figure \ref{HMEx} (left panel).  Wave 2, shown in Figure \ref{HMEx} (right panel), involves designing a set of three more runs over $ \setx_1 $, constructing another emulator over this region and calculating implausibility measures for each $ x \in \setx_1 $.  This second emulator is considerably more accurate than the observed error, thus $ \setx_2 \approx \bestset $, so the analysis can be stopped at this point as extra runs would do little to further reduce the non-implausible space. 


\subsection{Sequential History Matching of Observations}

\rev{Part of the novelty of our history matching approach involves dividing experimental observations into subsets and sequentially performing a history match on the model using each group of observations.}  Much scientific insight can be gained from performing a history match, \rev{however, using all output components simultaneously can mask which experiments are informative for certain aspects of the scientific system}.  

Breaking the data down into subsets and sequentially adding them to the history match is a novel approach \rev{which allows for further scientific insight.  Most prominently, it not only allows inferences to be made about the system quantities associated with the model input parameters, but also provides insight into the links between quantities associated with both the input and output.}   Note that \rev{adding model outputs sequentially in this way} is different from bringing outputs in sequentially due to emulator capability (step \rev{3} of the algorithm) \citep{GFBUA}. We will explore this in detail for the Arabidopsis model. 

\subsection{History Matching \rev{and} Bayesian \rev{Inference} \label{BHMMCMC} }

In this section we briefly discuss \rev{both} history matching \rev{and} the standard form of a full Bayesian analysis.

History matching is \rev{a computationally} efficient and practical approach to identifying if a model is consistent with observed data, and, if so, utilising the key uncertainties within the problem to identify where in the input space acceptable matches lie \citep{CSBS}.  In doing this, history matching attempts to \rev{answer some} of the main questions that a modeller may have.  \rev{A} full Bayesian framework requires full probabilistic specification of all \rev{uncertain} quantities, \rev{providing} a theoretically coherent method to obtain probabilistic answers \rev{to scientific} questions.  For example in the context of a direct simulator, as given by Equations (\ref{breakdown1}) and (\ref{breakdown2}), a posterior distribution for the location of the true best input $ \bestx $ is obtained. \rev{In comparision, as discussed in Section \ref{HM}, the non-implausible set resulting from a history match can be empty, thus contradicting \ian{the posited existence of $ \bestx $ and uncertainty specification associated with the ``best'' input approach}}. Problems \rev{may} arise if we do not believe that such a best input $ \bestx $ actually exists, since then a posterior over this input has little meaning.  \rev{In addition,} making full joint distributional specifications is challenging and frequently leads to approximations being made for mathematical convenience, \rev{which may call} into question the meaning of the resulting posterior.

Regardless of how prior distributions have been specified, performing the necessary calculations for a full Bayesian analysis is hard, thus requiring time consuming numerical schemes such as Markov Chain Monte Carlo (MCMC) \citep{brooks2011handbook}.  A major issue of numerical methods such as MCMC is that of convergence \citep{geyer2011introduction}.  Many model evaluations are required to thoroughly explore the multi-modal likelihoods over the entire input space.  Emulators can facilitate these large numbers of model evaluations at the cost of uncertainty \rev{\citep{BCCM, GPML, Higdon08a_calibration}}. However, since the likelihood function is constructed from all outputs of interest, we need to be able to emulate with sufficient accuracy all such outputs, including their possibly complex joint behaviour.  There may be erratically behaved outputs which are difficult to emulate, leading to emulators with large uncertainty or emulators which fail diagnostics. The likelihood, and hence the posterior, may be highly sensitive to these emulators, and hence be extremely non-robust.

Aside from the above concerns, the \rev{posterior distribution} of a Bayesian analysis (especially in high-dimensional models) often \rev{concentrates over a small subspace of the initial input domain of interest $ \setx_0 $.}
Obtaining a sufficiently accurate emulator over the whole input space will still require far too many model evaluations, and hence an iterative strategy such as history matching \rev{may be utilised}. History matching is designed to efficiently cut out the uninteresting parts of the input space, \rev{thus allowing} more accurate emulators \rev{to be constructed} over the region of interest $\bestset$, where the vast majority of the \rev{mass of the posterior distribution} should lie. A detailed Bayesian analysis can then be accurately performed within this much smaller volume of input space, \rev{where the full probabilistic specifications can now be considered more carefully}.
History matching can therefore be \rev{used} as a useful precursor to a full Bayesian analysis, or as a \rev{form of} analysis in its own right, particularly for modellers who do not wish to make the detailed specifications (or the corresponding robust analysis \citep{Berger:2000aa, berger1, watson2016}) that make the full Bayesian approach meaningful. \rev{There are other alternatives to history matching or a full Bayesian analysis, including the popular approach of Approximate Bayesian Computation (ABC) \citep{BSWT, ABCGE}, which could also be useful in this context.  In particular, \citet{AABCGP} incorporates history matching into an ABC paradigm.}  \rev{For further information about how history matching can fit into a Bayesian paradigm}, we refer the interested reader to the \ian{extended discussions in \cite{BUCSBM}, and for a direct comparison between history matching and ABC see \cite{ABCSBI}}.

\subsection{Application to Arabidopsis Model}

We now describe the relevant features of the hormonal crosstalk model as constructed by \citet{IPPHCARD}.

\subsubsection*{Description and Network}
The model represents the hormonal crosstalk of auxin, ethylene and cytokinin of Arabidopsis root development as a set of 18 differential equations, given in Table~\ref{DE}, which must be solved numerically.  The model takes an input vector of 45 rate parameters $ (k_1, k_{1a}, k_2, ...) $ and produces an output vector of 18 chemical concentrations $ ([Auxin], [X], [PLSp], ... ) $.  Note that, for simplicity, we refer to all components of the model, including hormones, proteins and mRNA, as ``chemicals''.  Experiments accumulated over many years have established certain relationships between some of the 18 concentrations.  For example, either manipulation of the PLS gene or exogenous application of IAA (a form of auxin), cytokinin or ACC (ethylene precursor) affects model outputs $ [Auxin] $, $ [CK] $, $ [ET] $ and $ [PIN] $.  We use initial conditions for the model, given in Table \ref{IOV}, that are consistent with \citet{MEAHCA, IPPHCARD}.

\begin{table}
\center
\small
\begin{eqnarray}
\frac{d[Auxin]}{dt} & = & \frac{k_{1a}}{\displaystyle 1 + \frac{[X]}{k_1}} + k_2 + k_{2a} \frac{[ET]}{\displaystyle 1 + \frac{[CK]}{k_{2b}}} \frac{[PLSp]}{k_{2c} + [PLSp]} \nonumber + \frac{V_{IAA}[IAA]}{Km_{IAA} + [IAA]} \nonumber \\ && \quad - \left( k_3 + \frac{k_{3a}[PIN1pm]}{k3auxin + [Auxin]} \right) [Auxin] \nonumber  \\
\frac{d[X]}{dt} & = & k_{16} - k_{16a}[CTR1^\ast] - k_{17}[X] \nonumber \\
\frac{d[PLSp]}{dt} & = & k_8[PLSm] - k_9[PLSp] \nonumber \\ 
\frac{d[Ra]}{dt} & = & -k_4 [Auxin] [Ra] + k_5 [Ra^\ast] \nonumber \\
\frac{d[Ra^\ast]}{dt} & = & k_4 [Auxin] [Ra] - k_5 [Ra^\ast] \nonumber \\
\frac{d[CK]}{dt} & = & \frac{k_{18a}}{\displaystyle 1 + \frac{[Auxin]}{k_{18}}} - k_{19} [CK] + \frac{V_{CK}[cytokinin]}{Km_{CK} + [cytokinin]} \nonumber \\
\frac{d[ET]}{dt} & = & k_{12} + k_{12a}[Auxin][CK] - k_{13}[ET] + \frac{V_{ACC}[ACC]}{Km_{ACC} + [ACC]} \nonumber \\
\frac{d[PLSm]}{dt} & = & \frac{k_6[Ra^\ast]}{\displaystyle 1 + \frac{[ET]}{k_{6a}}} - k_7[PLSm]  \nonumber \\
\frac{d[Re]}{dt} & = & k_{11}[Re^\ast][ET] - (k_{10} + k_{10a}[PLSp])[Re]   \nonumber \\
\frac{d[Re^\ast]}{dt} & = & -k_{11}[Re^\ast][ET] + (k_{10} + k_{10a}[PLSp])[Re] \nonumber \\
\frac{d[CTR1]}{dt} & = & -k_{14}[Re^\ast][CTR1] + k_{15}[CTR1^\ast] \nonumber \\
\frac{d[CTR1^\ast]}{dt} & = & k_{14}[Re^\ast][CTR1] - k_{15}[CTR1^\ast] \nonumber \\
\frac{d[PIN1m]}{dt} & = & \frac{k_{20a}}{k_{20b} + [CK]} [X] \frac{[Auxin]}{k_{20c} + [Auxin]} - k_{1_v21}[PIN1m] \nonumber \\
\frac{d[PIN1pi]}{dt} & = & k_{22a}[PIN1m] - k_{1_v23}[PIN1pi] - k_{1_v24}[PIN1pi] + \frac{k_{25a}[PIN1pm]}{\displaystyle 1 + \frac{[Auxin]}{k_{25b}}} \nonumber \\
\frac{d[PIN1pm]}{dt} & = & k_{1_v24}[PIN1pi] - \frac{k_{25a}[PIN1pm]}{\displaystyle 1 + \frac{[Auxin]}{k_{25b}}} \nonumber \\
\frac{d[IAA]}{dt} & = & 0 \nonumber \\
\frac{d[cytokinin]}{dt} & = & 0 \nonumber \\
\frac{d[ACC]}{dt} & = & 0 \nonumber 
\end{eqnarray} 
\normalsize
\caption[Arabdiopsis Equations]{Arabidopsis model differential equations. \label{DE}}
\end{table}

\begin{figure*}[h]
\center
\includegraphics[width=12cm]{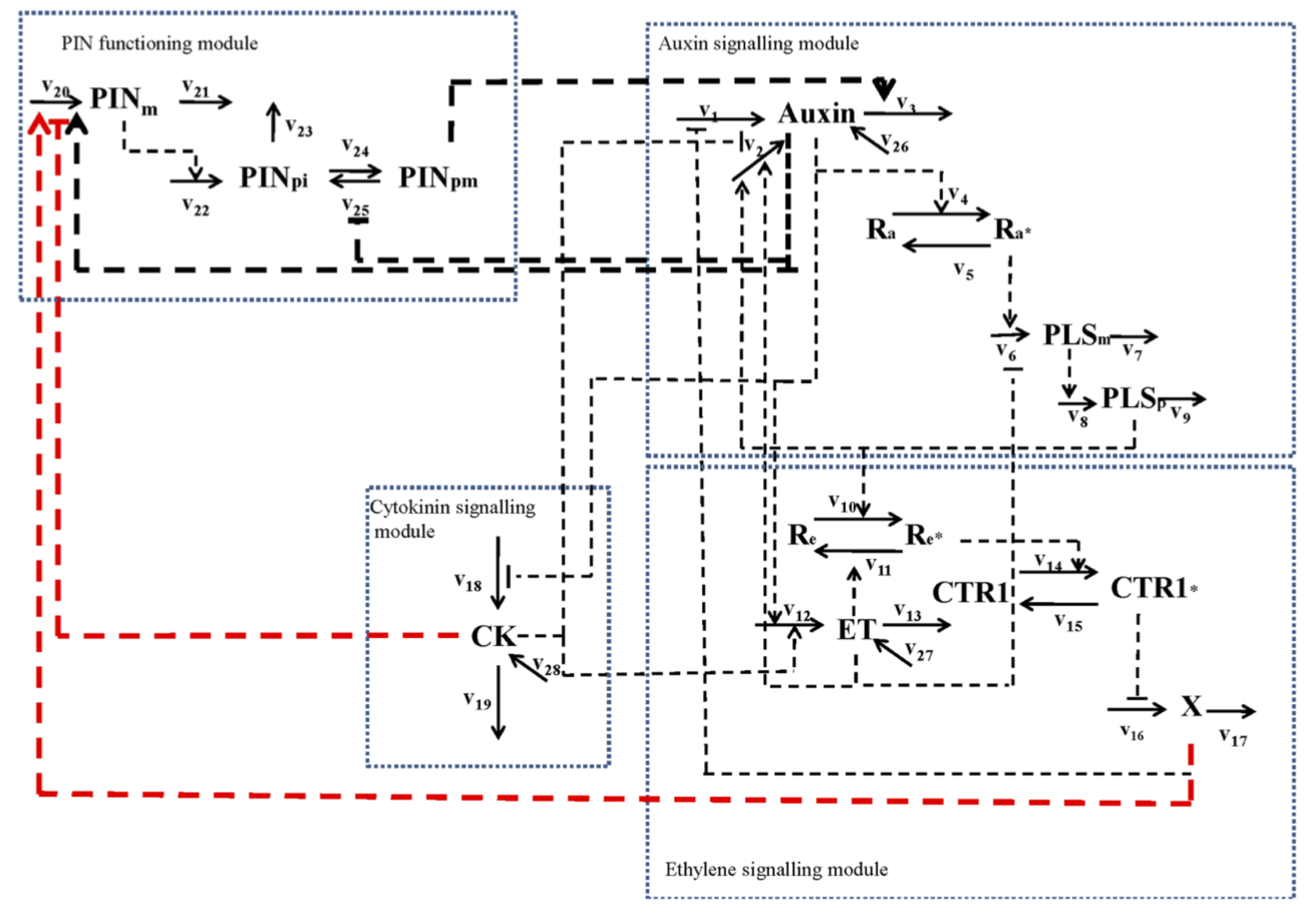}
\caption[Arabidopsis model network]{The Arabidopsis model network for the interaction of PIN, PLS and hormonal crosstalk.  The auxin, cytokinin and ethylene signalling modules correspond to the model of \citep{MEAHCA}.  The PIN functioning module is the additional interaction of the PIN proteins introduced in \citep{IPPHCARD}.  Solid arrows represent conversions whereas dotted arrows represent regulations.  The $ v_i $ represent reactions in the biological system and link to the rate parameters $ k_i $ on the right hand side of the differential equations in Table \ref{DE}. \label{AM2} }
\end{figure*}

The hormonal crosstalk network of auxin, cytokinin and ethylene for Arabidopsis root development is shown in Figure \ref{AM2}.  The auxin, cytokinin and ethylene signalling modules correspond to the model of \citet{MEAHCA}.  The PIN functioning module is the additional interaction of the PIN proteins introduced in \citet{IPPHCARD}.  Solid arrows represent conversions whereas dotted arrows represent regulations.  The $ v_i $ represent reactions in the biological system and link to the rate parameters $ k_i $ on the right hand side of the equations in Table \ref{DE}.  For full details of the model see \citet{IPPHCARD}.

\nomenclature{$ v_i $}{reactions in the model of \citet{IPPHCARD}}
\nomenclature{$ k_i $}{chemical reaction rates}

\begin{table}[ht] 
	\center
	\begin{tabular}{cc|cc}
		Output & Initial & Output & Initial \\ 
		& Concentration & & Concentration \\ \hline 
		$ Auxin $ & 0.1 & $ Re\ast $ & 0.3 \\
		 $ X $ & 0.1 & $ CTR1 $ & 0 \\ 
		 $ PLSp $ & 0.1 & $ CTR1\ast $ & 0.3 \\ 
		 $ Ra $ & 0 & $ PIN1m $ & 0 \\ 
		 $ Ra\ast $ & 1 & $ PIN1pi $ & 0 \\ 
		 $ CK $ & 0.1 & $ PIN1pm $ & 0 \\
		 $ ET $ & 0.1 & $ IAA $ & 0 or 1 \\ 
		 $ PLSm $ & 0.1 & $ cytokinin $ & 0 or 1 \\ 
		 $ Re $ & 0 & $ ACC $ & 0 or 1 \\
	\end{tabular}
	\caption[Model 1 Outputs]{The list of 18 original model outputs, along with their initial conditions.  The values of 0 or 1 for IAA, cytokinin and ACC correspond to no feeding or feeding of Auxin, Cytokinin or Ethylene respectively. See \citep{MEAHCA} and \citep{IPPHCARD} for details.  \label{IOV}}
\end{table}

\subsubsection*{Mutants and Feeding}

We will be interested in comparing the differences in chemical concentrations (corresponding to model outputs $ [Auxin], [CK], [ET], [PLSm] $ and $ [PIN] $) for different mutants (wild type (\emph{WT}), \emph{pls} mutant, PLS overexpressed transgenic (\emph{PLSox}), ethylene insensitive \emph{etr1}, double mutant \emph{plsetr1}) and feeding regimes (no feeding \emph{$ f_0 $}, feeding auxin \emph{$ f_a $}, feeding cytokinin \emph{$ f_c $}, feeding ethylene \emph{$ f_e $}, feeding any combination of these hormones $ f_af_c $, $ f_af_e $, $ f_cf_e $ and $ f_af_cf_e $) of Arabidopsis \citep{IPPHCARD}.  Note that wild type (\emph{WT}) refers to the typical plant occurring naturally in the wild that has not been mutated, however, we include this unmutated option in the list of mutants for notational convenience.  Note that for simplicity of terminology, exogenous application of IAA, cytokinin or ACC is referred to as ``feeding auxin, cytokinin or ethylene" respectively.

In the model, mutant type is controlled by altering the parameters representing the expression of the two genes \emph{PLS} and \emph{ETR1}.  Input rate parameter $ k_6 $ controls the amount to which PLS is suppressed, hence \emph{pls} is represented by setting $ k_6 = 0 $ and \emph{PLSox} is represented by increasing the size of $ k_6 $ to a value greater than that of the wild type plant.  Input rate parameter $ k_{11} $ represents the rate of conversion of the active form of the ethylene receptor to its inactive form.  The ethylene insensitive \emph{etr1} mutant is represented by decreasing the size of $  k_{11} $ to a much smaller value than that of wild type.  \emph{plsetr1} is represented by both setting $ k_6 = 0 $ and $ k_{11} $ to its much decreased value.  
Feeding regime is represented by the initial conditions of certain outputs.  $ [IAA] $, $ [cytokinin] $ and $ [ACC] $ take initial condition values 0 or 1, as indicated in Table \ref{IOV}, depending on whether or not the respective chemical auxin, cytokinin or ethylene has been fed.

\subsubsection*{Model Structure and The Inputs}

Model structure sometimes restricts what we are able to learn about certain parameter relationships.  \rev{For example, in this case, there is a constraint that $ k_{16}/k_{16a} = 0.3 $, which ensures that the term $ k_{16} - k_{16a}[CTR1^\ast] $ in the $ d[X]/dt $ equation is non-negative, thus effectively removing an input from the set of rate parameters in the equations in Table \ref{DE}. In principle, given sufficient runs, history matching should discover such restrictions in the model (for example, \ian{this restriction was identified for a simpler model via history matching in \cite{BUCSBM}}), but the ability to identify these restrictions before we start will make the process more efficient.}

\rev{In addition to this restriction,} we are only interested in comparing the model output to data at equilibrium,  thus allowing a substantial dimensional reduction of the input space. At equilibrium, the derivatives on the left hand side of the model equations given in Table \ref{DE} 
will equal zero, and hence the right hand side can be rearranged in terms of one less parameter~\citep{BUCSBM}.
For this reason, measurements of outputs of this system will only allow us to learn about certain ratios of the input rate parameters to one another.  For example, the equation for $d[PLSp]/dt$ becomes
\begin{eqnarray} 
\frac{d[PLSp]}{dt}  \;\; =\;\; 0 &=&  k_8[PLSm] - k_9[PLSp]   \\
\Rightarrow \quad\quad 0 & =&  \frac{k_8}{k_9}[PLSm] - [PLSp] 
\end{eqnarray}
which only depends on the ratio $ k_8/k_9 $.

Another restriction arises from the fact that the initial conditions for the feeding chemicals $ [IAA] $, $ [cytokinin] $ and $ [ACC] $ can only take the values 0 or 1 and then remain constant.  This is because, although the expressions $ \frac{V_{IAA}[IAA]}{Km_{IAA} + [IAA]} $, $ \frac{V_{CK}[cytokinin]}{Km_{CK} + [cytokinin]} $ and $ \frac{V_{ACC}[ACC]}{Km_{ACC} + [ACC]} $ respectively in the equations for $ [Auxin] $, $ [CK] $ and $ [ET] $ take the specific form following the biological mechanism, they can only be learnt about as a whole, essentially comparing the case of a constant reservoir of chemical being available for uptake into the plant with the case of no feeding at all.  Feeding of IAA, cytokinin and ACC with any concentration can be rescaled to $ [IAA]=1 $, $[cytokinin]=1 $, and $ [ACC]=1 $ by adjusting the parameters $ V_{IAA} $, $ V_{CK} $, and $ V_{ACC} $ in each equation respectively. Note that specific equations for the rate of change of the feeding chemicals may allow more insight into the effects of feeding if deemed biologically relevant.  

\rev{Following the previous section, we let $ k_{6w} $ and $ k_{11w} $ represent the values that $ k_6 $ and $ k_{11} $ respectively should take for wild type.  We let the two additional parameters $ k_{6m} > 1 $ and $ k_{11m} << 1 $ represent the values these parameters should be multiplied by in order to obtain the corresponding model run for the \emph{PLSox} and \emph{etr1} mutants respectively, that is with $ k_6 = k_{6m}k_{6w} $ and $ k_{11} = k_{11m}k_{11w} $.  Doing this allows exploration of a reasonable class of representations of these mutants using independent parameters.  }

Finally, since we consider ranges of rate parameters and rate parameter ratios which are always positive, many spanning many orders of magnitude, we choose to convert \rev{them} to a log scale.  We therefore define the reduced 31 dimensional vector of input parameters for the model to be:
\begin{equation}
x = \log(k_1, k_{1a}/k_2, k_{2a}/k_2, ..., k_{11m}) 
\end{equation}
as given in \rev{the left hand column of} Table~\ref{RR}. 

In order to perform a full analysis on the model, we introduce a parameter $ \lambda = V_i/V_m $ to represent the ratio of the cytosolic volume $ V_i $ to the volume of the cell wall $ V_m $.  Full details of why we introduce this parameter are included in Appendix \ref{EPlambda}.  For a typical cell, we fixed $ \lambda = 6 $ and considered that a reasonable range of possible values for $ \lambda $ was $ [2,16] $ for a plant root cell.

\begin{table}[p]
	\center
	\begin{tabular}{|c|c|c|c|} \hline
		Input Rate & Initial  & Minimum & Maximum \\ 
		Parameter (Ratio) & Value & & \\ \hline
		$ k_1 $ & 1 & 0.1 & 10 \\
		$ k_{1a}/k_2 $ & 5 & 0.5 & 50 \\ 
		$ k_{2a}/k_2 $ & 14 & 1.4 & 140  \\
		$ k_{2b} $ & 1 & 0.1 & 10 \\
		$ k_{2c} $ & 0.01 & 0.000001 & 0.1 \\
		$ k_3/k_2 $ & 10 & 1 & 100 \\
		$ k_{3a}/k_2 $ & 2.25 & 0.225 & 22.5 \\
		$ k_{3auxin} $ & 10 & 1 & 100 \\
		$ k_5/k_4 $ & 1 & 0.1 & 10 \\	 
		$ k_{6a} $ & 0.2 & 0.002 & 2000  \\
		$ k_{6w}/k_7 $ & 0.3 & 0.03 & 3 \\
		$ k_9/k_8 $ & 1 & 0.1 & 10 \\
		$ k_{10a}/k_{10} $ & 16600 & 166 & 16600  \\
		$ k_{11}/k_{10} $ & 16600 & 16.6 & 166000  \\
		$ k_{12a}/k_{12} $ & 1 & 0.1 & 10 \\
		$ k_{13}/k_{12} $ & 10 & 1 & 1000  \\
		$ k_{15}/k_{14} $ & 0.0283 & 0.000283 & 0.283 \\
		$ k_{17}/k_{16a} $ & 0.1 & 0.01 & 1 \\
		$ k_{19}/k_{18a} $ & 1 & 0.1 & 10  \\
		$ k_{18} $ & 0.1 & 0.01 & 10 \\
		$ k_{20a}/k_{1_v21} $ & 0.8 & 0.08 & 8  \\
		$ k_{20b} $ & 1 & 0.1 & 10 \\
		$ k_{20c} $ & 0.3 & 0.03 & 3  \\
		$ k_{22a}/k_{1_v23} $ & 1.35 & 0.135 & 13.5  \\
		$ k_{25a}/k_{1_v24} $ & 0.1 & 0.01 & 1  \\
		$ k_{25b} $ & 1 & 0.1 & 10  \\
		$ V_{IAA}/k_2(Km_{IAA}+1) $ & 2.27 & 0.05 & 50 \\
		$ V_{CK}/k_{18a}(Km_{CK}+1) $ & 0.45 & 0.01 & 1  \\
		$ V_{ACC}/k_{12}(Km_{ACC}+1) $ & 4.55 & 0.1 & 100  \\
		$ k_{6m} $ & 1.5 & 1 & 4 \\
		$ k_{11m} $ & 0.006 & 0.001 & 0.1 \\ \hline
	\end{tabular}
	\caption[Input Ranges for Arabidopsis PIN model]{A table of parameter ranges (which were converted to $ [-1,1] $ for the analysis).  These define the initial search region $\setx_0$. \label{RR}}
\end{table} 


\subsection{Eliciting the Necessary Information for History Matching}

To perform a history match, we need to understand how real-world observations relate to model outputs, thus aiding the specification of observed values $ z_i $, model discrepancy terms $ \sigma_{\epsilon_i} $ and measurement error terms $ \sigma_{e_i} $.  History matching is a versatile technique which can deal with observations of varying quality, such as we have for the Arabidopsis model.

\subsubsection*{Relating Observations To Model Outputs}

Each Arabidopsis model output relating to a biological experiment can be represented by: $$ h_{j,m,a}(x,t) $$ where: 
\begin{eqnarray} 
j & \in & \{[Auxin], [PLSm], [CK], [ET], [PIN]\} \nonumber \\  
m & \in & \{WT, pls, PLSox, etr1, pls etr1\} \nonumber  \\ 
a & \in & \{f_0, f_a, f_c, f_e, f_af_c, f_af_e, f_cf_e, f_af_cf_e \}  \nonumber 
\end{eqnarray}
Here, the subscript $ j $ indexes the measurable chemical, $ m $ indexes the plant type and $ a $ indexes the feeding action, where $ f_0 $ indicates no feeding and $ f_a $, $ f_c $ and $ f_e $ indicate the feeding of auxin, cytokinin and/or ethylene respectively, for a particular set-up of the general model $ h $ (the Arabidopsis model equations given in Table \ref{DE}).  The vector $ x $ represents the vector of rate parameter ratios and $ t $ represents time.  There are 200 possible experiments given by the possible combinations of $ j $, $ m $ and $ a $.

The average PIN concentration in both the cytosol and the cell wall is calculated as follows:
\begin{equation} 
[PIN] = \frac{[PIN1pm] + \lambda [PIN1pi]}{1+\lambda} 
\end{equation} 

We collected the results of a subset of 32 of the possible experiments from a variety of experiments in the literature (see \citet{MEAHCA, IPPHCARD} and references therein for details).  30 of these observations are measures of the trend of the concentration of a chemical for one experimental condition relative to another experimental condition  (usually chosen to be wild type).  We therefore need our outputs of interest to be ratios of the outputs of our model $ h $ with different experimental subscript settings.  We choose to work with log model outputs since these will be more robust and allow multiplicative error statements.  Since we only consider model outputs to be meaningful at equilibrium, that is as $ t \rightarrow \infty $, we therefore, following \citet{BUCSBM}, define the main outputs of interest to be:
\begin{equation} 
f_i(x) = \lim_{t \rightarrow \infty} \log \left\{ \frac{h_{j, m_2, a_2}(x,t)}{h_{j, m_1, a_1}(x,t)} \right\}
\end{equation} 
where the subscript $ i $ indexes the combinations of $ \{j, m_1, a_1, m_2, a_2 \} $ that were actually measured.  This function $ f_i(x) $ will be directly compared to the observed trends.  All but one of the trends were relative to wild type with no feeding, with the exception being the ratio of auxin concentration in the \emph{pls} mutant fed ethylene to the \emph{pls} mutant without feeding.  The remaining two observations are non-ratio wild type measurements of the chemicals $ [Auxin] $ and $ [CK] $.  The outputs of interest for these observations are given as $ \lim_{t \rightarrow \infty} \log \{h_{[Auxin], WT, f_o}(x,t) \} $ and $ \\ \lim_{t \rightarrow \infty} \log \{h_{[CK], WT, f_o}(x,t) \} $ respectively.  Including these experiments within the history match ensures that acceptable matches will not have unrealistic concentrations of auxin and cytokinin.

The full list of 32 outputs is given in the left hand column of Table \ref{OR}.  These are notated in the form:  
\begin{equation}
 mutant(\mbox{if not wild type})\_feeding(\mbox{if any})\_chemical 
\end{equation}
and are assumed to be ratios relative to wild type with no feeding unless otherwise specified.  NR indicates that an output is not a ratio.  For example, $ f_e\_CK $ indicates the cytokinin concentration ratio of wild type fed ethylene relative to wild type no feeding, and $ PLSox\_ET $ represents the ethylene concentration ratio of the POLARIS overexpressed mutant relative to wild type.

We sequentially history match the Arabidopsis model to these experimental observations in 3 phases $ A $, $ B $ and $ C $, with the group to which each experiment belongs presented in Table \ref{OR}.  We will history match the Dataset $ A $ observations to obtain a non-implausible set $ \setx_A $.  Additional insight will be gained by further history matching to Dataset $ B $ to obtain $ \setx_B $, and then finally history matching to Dataset $ C $.  Dataset $ B $ contains the outputs involving the feeding of ethylene.  History matching this group separately provides insight into how the inputs of the model are constrained based on physical observations of a plant having been fed ethylene relative to its wild type counterpart.  Dataset $ C $ contains the outputs involving the measurement of $ PLSm $, thus demonstrating how useful observing the effects of the POLARIS gene function were for gaining increased understanding about the model and its rate parameters.

\begin{table*}[p] 
	\center
	\small
		\begin{tabular}{|c|c|c|c|c|c|} \hline
	 & & Minimum & Maximum & Minimum & Maximum \\ 
	Experiment & Dataset & Log Ratio & Log Ratio & Ratio & Ratio  \\ 
	& & Value &  Value & Value & Value \\ \hline
$ WT\_Auxin $ (NR) & A & -3.772 & 0.833 & 0.023 & 2.3 \\
$ pls\_Auxin $ & A & -1.531 & 0.366 & 0.216 & 1.442 \\
$ PLSox\_Auxin $ & A & -0.576 & 0.708 & 0.562 & 2.031 \\
$ etr1\_Auxin^\star $ & A & 0.182 & 2.303 & 1.2 & 10 \\
$ plsetr1\_Auxin $ & A & -0.792 & 0.600 & 0.453 & 1.823 \\
$ f_a \_Auxin^\star $ & A & 0.182 & 2.303 & 1.2 & 10 \\
$ f_c\_Auxin $ & A & -2.303 & 1.099 & 0.1 & 3 \\
$ f_e\_Auxin^\star  $ & B & 0.182 & 2.303 & 1.2 & 10 \\
$ pls\_f_e\_Auxin/pls\_Auxin $ & B & -1.204 & -0.010 & 0.3 & 0.99 \\
$ WT\_CK $ (NR) & A & -3.730 & 0.875 & 0.024 & 2.4 \\
$ pls\_CK $ & A & 0.049 & 1.253 & 1.05 & 3.5 \\
$ PLSox\_CK^\star $ & A & -2.303 & -0.182 & 0.1 & 0.834 \\
$ f_a\_CK^\star $ & A & -2.303 & -0.182 & 0.1 & 0.834 \\
$ f_c\_CK^\star $ & A &  0.182 & 2.303 & 1.2 & 10 \\
$ f_e\_CK^\star $ & B & -2.303 & -0.182 & 0.1 & 0.834 \\
$ pls\_ET^\star $ & A & -0.342 & 0.336 & 0.71 & 1.4 \\
$ PLSox\_ET^\star $ & A &  -0.342 & 0.336 & 0.71 & 1.4 \\
$ f_a\_ET^\star $ & A &  0.182 & 2.303 & 1.2 & 10 \\
$ f_c\_ET^\star $ & A &  0.182 &  2.303 & 1.2 & 10 \\ 
$ f_e\_ET^\star $ & B &  0.182 & 2.303 & 1.2 & 10 \\
$ f_a\_PLSm^\star $ & C &  0.182 & 2.303 & 1.2 & 10 \\
$ f_c\_PLSm^\star $ & C & -2.303 & -0.182 & 0.1 & 0.834 \\
$ f_e\_PLSm^\star $ & C & -2.303 & -0.182 & 0.1 & 0.834 \\
$ f_af_c\_PLSm $ & C &  -0.554 & 3.449 & 0.575 & 31.482 \\
$ f_af_e\_PLSm $ & C & 0.207 & 3.315 & 1.23 & 27.528 \\
$ pls\_PIN $ & A & -0.650 & 1.007 & 0.522 & 2.738 \\
$ PLSox\_PIN $ & A & -1.629 & 0.456 & 0.196 & 1.578 \\
$ etr1\_PIN $ & A & -1.892 & 0.182 & 0.151 & 1.199 \\
$ plsetr1\_PIN $ & A & -1.175 & 0.613 & 0.309 & 1.846 \\
$ f_a\_PIN^\star $ & A &  0.182 & 2.303 & 1.2 & 10 \\
$ f_c\_PIN^\star $ & A & -2.303 & -0.182 & 0.1 & 0.834 \\
$ f_e\_PIN $ & B & -0.730 & 0.893 & 0.482 & 2.443 \\ \hline
		\end{tabular}
		\normalsize
	\caption[Emulator Output Ranges]{A table showing the natural ranges and logarithmic ranges of simulator output values that would be accepted at implausibility cutoff 3.  Column 2 shows which of the three Datasets each output belongs to. These \rev{outputs} are notated in the form $ mutant(\mbox{if not wild type})\_feeding(\mbox{if any})\_chemical $ and are assumed to be ratios \rev{of the output for the specified mutant} relative to \rev{that for} wild type with no feeding unless otherwise specified.  NR indicates that an output is not a ratio, and * indicates that the data for that experiment was a general trend.   \label{OR}}
\end{table*}

\subsubsection*{Observed Value, Model Discrepancy and Measurement Error}

Although some of our collected measurement values were estimated values of a trend or ratio, many of the measurements were only general trend directions or estimated ranges for the ratio value, given with various degrees of accuracy \citep{IPPHCARD}.  We therefore use a level of modelling appropriate to the nature of the data to propose order of magnitude estimators for $ \obs_i $, $ \sigma_{\epsilon_i} $ and $ \sigma_{e_i} $  that are consistent with the observed trends and expert judgement concerning the accuracy of the model and the relevant experiments.  Doing this demonstrates that we can apply our history matching approach to vague, qualitative data, whilst demonstrating the increased power of this analysis were we to have more accurate quantitative data for all the experiments.  

A general trend of ``Up'', ``Down'' or ``No Change'' was collected for 17 of the experiments, these being indicated by an asterix in Table \ref{OR}. Following the conservative procedure given in \citep{BUCSBM}, we specify $ \obs_i = 1.24, -1.24  \mbox{ and }  0 $ and $ \sigma_{c_i} = 0.35, 0.35 \mbox{ and } 0.061$ for the ``Up'', ``Down'' and ``No Change'' trends respectively, where  $ \sigma_{c_i} $ represents the combined model discrepancy and measurement error, that is $ \sigma_{c_i} = \sqrt{\sigma_{\md_i}^2 + \sigma_{\me_i}^2} $.  These combined specifications have been chosen such that $ z_i \pm 3 \sigma_{c_i} $ represents a 20\% to ten fold increase for the ``Up'' trends, a 20\% to ten fold decrease for the ``Down'' trends, and a 40\% decrease to 40\% increase for the ``No Change'' trends.  To avoid confusion, we here define a 20\% decrease to imply that a 20\% increase on the decreased value returns the original value.  This specification conservatively captures the main features of the trend data, although more in-depth specification could be made if quantitative measurements were available across these outputs.  We specify $ z_i $ to be in the middle of the logged ratio range.  In this work we considered that the deficiencies in the model would be of a similar order of magnitude to the observed errors on the data.  We therefore specify both model discrepancy and measurement error to be of equal size and satisfy the ratio intervals above.

For the remaining cases, the observed values $ z_i $, model discrepancies $ \sigma_{\md_i} $ and measurement errors $ \sigma_{\me_i} $ were chosen using a more in-depth expert assessment of the accuracy of the relevant trend measurements and their links to the model (see \citet{MEAHCA, IPPHCARD} and references therein for details).  Since we will use a maximum implausibility threshold of $c=3$ by appealing to Pukelsheim's 3 sigma rule \citep{3SR} when working with the simulator runs, it is most appropriate to specify the logged ranges of $ z_i \pm 3 \sigma_{c_i} $, as these are the ranges which if a simulator run falls outside it will be classed as implausible.  These ranges are specified in Table \ref{OR} in both logged and not logged form.

\subsubsection*{Input Ratio Ranges}

The initial ranges of values for the \rev{31} input parameters were chosen based on those in the literature \citep{MEAHCA} and further analysis of the model \citep{IPPHCARD}, and are shown in Table \ref{RR}.  Many of the input ranges were chosen to cover an order of magnitude either side of the single satisfactory input parameter setting found in \citet{MEAHCA}.  Some parameters of particular interest were subsequently increased to allow a wider exploration of the input parameter space.  This gave us a large initial input space $ \setx_0 $ which was thought to be suitable for our purposes.  The logged ratio ranges were all converted to the range $ [-1,1] $ prior to analysis.  

We now apply the technique of sequential history matching using Bayes linear emulation to the Arabidopsis model \citep{IPPHCARD}.  Analysis of the results, after history matching to each of Datasets $ A $, $ B $ and $ C $, will involve consideration of the following:
\begin{itemize}
\item The volume reduction of the non-implausible input space \citep{BUCSBM}.
\item Input plots of the non-implausible space \citep{BUCSBM}.
\item The variance resolution of individual inputs and groups of inputs.
\item Output plots of the non-implausible space \citep{BUCSBM}.
\item The degree to which each output was informative for learning about each input.
\end{itemize}


\section{Results}


\subsection{Insights From Initial Simulator Runs}

\begin{figure*}[h]
\center
\includegraphics[width=12cm]{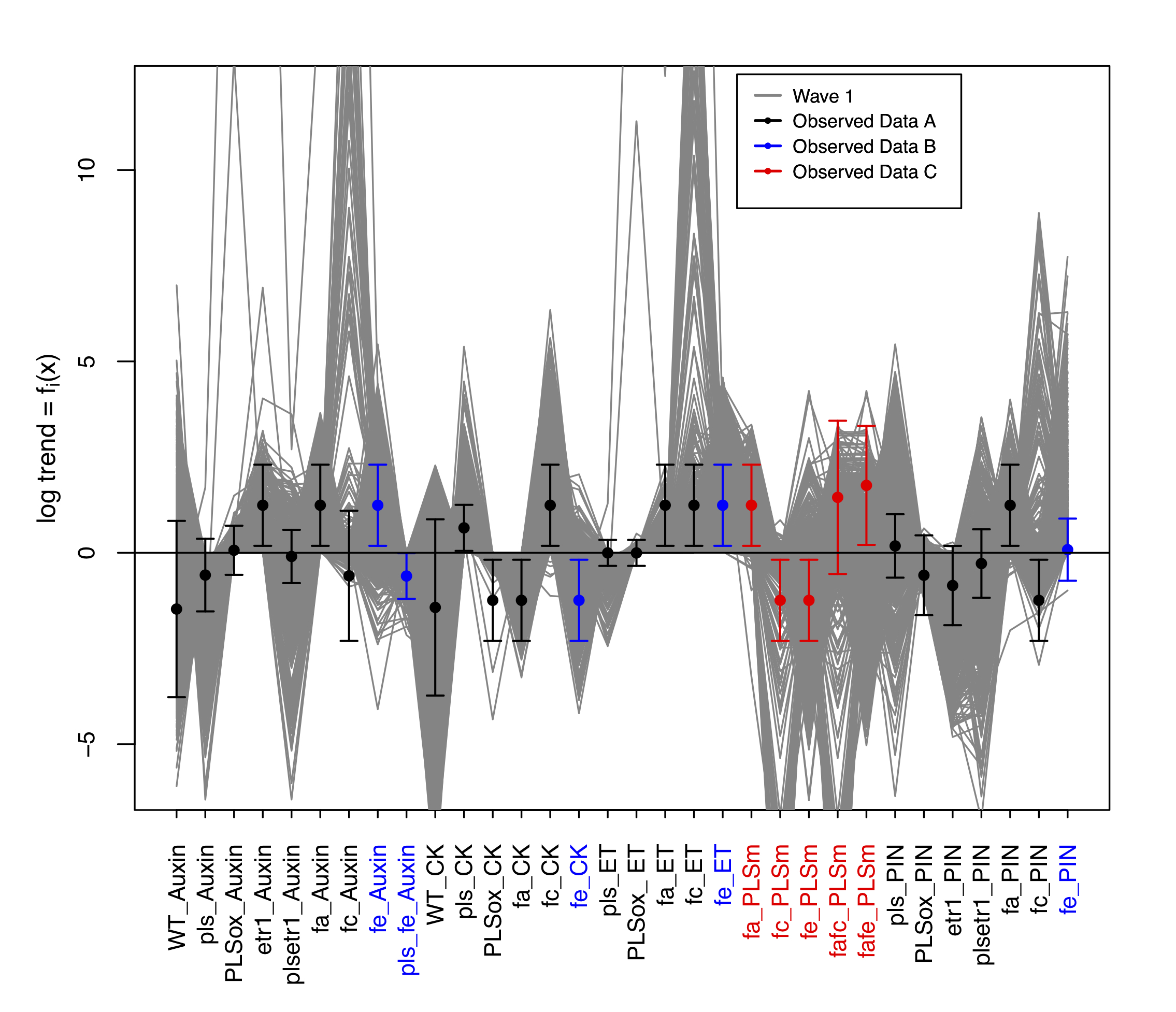}
\caption[1D Output Plot of Initial Simulator Runs]{2000 wave 1 output runs $ f_i(x) $ for all 32 outputs considered.  The targets for the history match, as given by the intervals $ z_i \pm 3 \sigma_{c_i} $ and the ranges in Table \ref{OR}, are shown as vertical error bars.  Black error bars represent Dataset $ A $ outputs, blue error bars represent Dataset $ B $ outputs and red error bars represent Dataset $ C $ outputs.  The horizontal black line at zero represents zero trend.   \label{IW1R} }
\end{figure*}

A wave 1 set of 2000 training runs were designed using a maximin Latin hypercube design over the initial input space $ \setx_0 $.  Figure \ref{IW1R} shows the wave 1 output runs $ f_i(x) $ for all 32 outputs considered.  The targets for the history match, as given by the intervals $ z_i \pm 3 \sigma_{c_i} $ and the ranges in Table \ref{OR}, are shown as vertical error bars.  Black error bars represent Dataset $ A $ outputs, blue error bars represent Dataset $ B $ outputs and red error bars represent Dataset $ C $ outputs.  \rev{Note that the measurements for Datasets $ B $ and $ C $ are shown for illustrative purposes, and in general may have been obtained at a later point in time to when the wave 1 model runs were performed.  In this case, however, the model would still produce a value for all output components (both those with and without corresponding physical measurements)}.

Figure \ref{IW1R} gives substantial insight into the general behaviour of the model over the initial input space $\setx_0$, for example informing us about model outputs that can take extreme values, for example, $ f_c\_Auxin $, $ f_c\_{ET} $ and $ f_af_c\_PLSm $. More importantly, the runs also inform us as to the class of possible observed data sets that the model could have matched, and hence gives insight into the model's flexibility. There exist outputs with constrained ranges.  In particular, many outputs seem to be constrained to being either positive or negative, 
for example, the logged trend for $ pls\_CK $ must be positive and that of $ PLSox\_CK $ must be negative. If such constrained outputs, which are consequences of the biological structure of the model, are found to be consistent with observations, this provides (partial) evidence for the model's validity. Conversely, we may be concerned about an overly flexible model that was capable of reproducing any combination of positive or negative observed data values for outputs in this dataset.  Specifically, we should doubt claims that such a model has been validated by comparison to this data, as it would have inevitably matched any possible data values and hence arguably may not contain much inherent biological structure at all. 

There are some outputs for which the majority of the wave 1 runs already go through the corresponding error bars, for example $ PLSox\_Auxin $ and $ PLSox\_ET $.  This is an indication that these outputs did not help much to constrain the input space.  Despite this, none of the wave 1 runs pass through all of the target intervals of the outputs in Dataset $ A $ simultaneously, thus already suggesting that the volume of the final non-implausible space would be small or indeed zero.

\subsection{History Matching The Model}

We outline the general decisions required to perform the history match.  Several packages are available that perform standard Gaussian Process emulation, possibly with Automatic Relevance Determination \citep{GPR, MCIGPM}, for example the BACCO \citep{Hankin:2005aa} and GPfit \citep{JSSv064i12} packages in R (R Core Team) or GPy \citeyearpar{gpy2014} for Python, which may be used as an alternative to the emulators we describe here. Emulators accurate enough to reduce the size of the non-implausible space at each wave to some degree are sufficient \rev{for our purposes}.  When constructing emulators, we decided to put more detail into the mean function, but incorporate more complicated structures for the residual process at each wave, thus sequentially increasing the complexity of the emulators at each wave.  We provide a summary of the choices made in the history match at each wave in Table \ref{ES}, including the dataset history matched to (column 2), the number of design runs (column 3), the implausibility cut-off thresholds (columns 4-6) and the emulation strategy (column 7), each of which is discussed in more detail below.

The amount of space that was cut out after each wave is shown in Table \ref{ISR}.  We let $ V(\setx_i )$ represent the volume of the non-implausible space after wave $ i $, as judged by the emulators, and $ V(\setx_G) $ represent the volume of the space with acceptable matches to the observed data in Dataset $ G $, as judged using actual model runs (hence without emulator error).  Then columns 2 and 3 give the proportion of the previous wave and initial non-implausible spaces respectively still classed as non-implausible, and columns 5 and 6 give the proportion of the wave $ i $ and initial non-implausible spaces giving rise to actual acceptable matches to the data in Dataset $ G $.  The proportion of space cut out at each wave is influential for deciding the number of waves and emulator technique at each wave.  In addition, Table \ref{ISR} presents the radical space reduction obtained by performing the history match.  A proportion of $ 6.1 \times 10^{-7} $ of the original space was still considered non-implausible after history matching to Dataset $ A $.  A proportion of only $ 8.5 \times 10^{-10} $ of the original space was still considered non-implausible after history matching to Datasets $ A $ and $ B $, thus the 5 trends in Dataset $ B $, for exogenous application of ACC, facilitated an additional reduction of 3 orders of magnitude.  After all experimental observations had been matched to, the non-implausible space had been reduced to a proportion of $ 7.2 \times 10^{-12} $ of the original space, thus the 5 trends in Dataset $ C $, for measurement of POLARIS gene expression, refocused the set by another 2 orders of magnitude.  Such small proportions of the original space being classed as non-implausible means that acceptable runs within these spaces would likely be missed by more ad-hoc parameter searching methods of analysis.
%

\begin{table*}[ht]
	\center
	\small
	\begin{tabular}{|c|c|c|ccc|c|} \hline
		Wave ($ i $) & Dataset ($ D $) & Runs & $ I_M^{cut} $ & $ I_{2M}^{cut} $ & $ I_{3M}^{cut} $ & Emulation Strategy \\ \hline
		1 & $ A $ & 2000 & - & 3 & 2.9 & Linear models \\
		&&&&&& \\
		2 & $ A $ & 2000 & 3 & 2.9 & - & Linear models \\
		&&&&&& \\
		3, 4 & $ A $ & 2000 & 3 & 2.8 & - & Linear models \\
		&&&&&& \\
		5 & $ A $ & 2000 & 3 & 2.8 & - & Single fixed  \\
		&&&&&& correlation length \\
		6, 7 & $ A $ & 2000 & 3 & 2.8 & - & Several correlation  \\
		&&&&&& lengths per output \\
		8, 9 & $ A, B $ & 2000 & 3 & 2.9 & - & Linear models for  \\
		&&&&&& Dataset B outputs only \\
		10 & $ A, B $ & 2000 & 3 & 2.9 & - & Single fixed \\
		&&&&&& correlation length \\
		11 & $ A, B $ & 3500 & 3 & 2.9 & - & Several correlation\\
		&&&&&& lengths per output \\
		12 & $ A, B, C $ & 2000 & 3 & 2.9 & - & Single fixed  \\
		&&&&&& correlation length \\
		13 & $ A, B, C $ & 3500 & 3 & 2.9 & - & Several correlation \\ 
		&&&&&& lengths per output \\ \hline
\end{tabular}
	\caption[Input Space Reduction Per Wave]{A summary of the wave-by-wave emulation strategy.  Column 1: wave number.  Column 2: Datasets history matched at wave $ i $.  Column 3: Number of model runs used to construct the emulator.  Columns 4-6: Cutoff thresholds used at each wave for each of the implausibility critieria.  Column 7: Emulation strategy for wave $ i $. \label{ES}}
\end{table*}

\begin{table*}[ht]
	\center
	\begin{tabular}{|c|c|c|c|c|c|} \hline
		Wave ($ i $) & $ \frac{V(\setx_i)}{V(\setx_{i-1})} $ & $ \frac{V(\setx_i)}{V(\setx_0)} $ & Dataset ($ G $) &  $ \frac{V(\setx_G)}{V(\setx_i)} $ & $ \frac{V(\setx_G)}{V(\setx_0)} $ \\ \hline
		1 &  0.45 	& $ 4.5 \times 10^{-1} $ &  &  & \\ 
		2 &  0.12 	& $ 5.4 \times 10^{-2} $ &  &  & \\
		3 &  0.035 	& $ 1.9 \times 10^{-3} $ &  &  & \\
		4 &  0.25 	& $ 4.7 \times 10^{-4} $ &  &  & \\
		5 &  0.12 	& $ 5.7 \times 10^{-5} $ &  &  & \\
		6 &  0.15 	& $ 8.5 \times 10^{-6} $ &  &  & \\
		7 &  0.55 	& $ 4.7 \times 10^{-6} $ & $ A $ & 0.13 & $ 6.1 \times 10^{-7} $ \\
		8 &  0.25 	& $ 1.2 \times 10^{-6} $ &  &  & \\
		9 &  0.11 	& $ 1.3 \times 10^{-7} $ &  &  & \\
		10 & 0.55 	& $ 7.1 \times 10^{-8} $ &  &  & \\
		11 & 0.15 	& $ 1.1 \times 10^{-8} $ & $ A, B $ & 0.08 & $ 8.5 \times 10^{-10} $ \\
		12 &  0.1  	& $ 1.1 \times 10^{-9} $ &  &  & \\
		13 &  0.45 	& $ 4.8 \times 10^{-10} $ & $ A, B, C $ & 0.015 & $ 7.2 \times 10^{-12} $ \\ \hline
\end{tabular}
	\caption[Input Space Reduction Per Wave]{A summary of the space cut out by the 13 waves of emulation and additional space cut out by the simulators for each dataset.  Column 2: proportion of previous wave's non-implausible space still classed as non-implausible.  Column 3: proportion of original space still classed as non-implausible.  Column 5: proportion of wave $ i $ non-implausible space giving rise to acceptable matches to the data in Dataset $ G $ using simulations.  Column 6: proportion of original space giving rise to acceptable matches to the data in Dataset $ G $ using simulations. \label{ISR}}
\end{table*}

Linear model emulators with uncorrelated residual processes were used in the initial waves since they are very cheap to evaluate, substantially more so even than emulators involving a correlated residual process, which may only be slightly more accurate \citep{HMHIV}.  \rev{For these emulators, we estimated the value of $ \sigma_{u_i}^2 $ to be the estimated variance parameter from the linear model fit.}  As the amount of space being classed as implausible at each wave started to drop, we introduced emulators with a Gaussian correlation residual process.  \rev{There are various methods in the literature for assessing correlation length parameters, as explained in Section \ref{BLE}.}  Some of the methods in the literature for picking the correlation lengths $ \theta $ \rev{and variance parameter $ \sigma_{u_i}^2 $} tend to be computationally intensive and their result highly sensitive to the sample of simulator runs \citep{PEPUG, PEGPE}.  The choice was therefore made, at wave 5, to use a single correlation length parameter value of $ \theta = 2 $ for all input-output combinations, \rev{and to fit $ \sigma_{u_i}^2 $ using the corresponding linear model fit, these choices} being checked using emulator diagnostics.  \rev{The motivation for this choice of correlation length parameter was made by appealing to the heuristic argument made in \citet{GFBUA} that the regression residuals may be derived from a polynomial of order one higher than the fitted polynomials, the alteration in the chosen value taking into account the higher dimensionality of the input space.}

At wave 6 the complexity of the residual process was increased still further by splitting the active inputs $x_{A_i}$ for each output emulator into five groups based on similar strength of effect, and using maximum likelihood to fit the same correlation length to all inputs in each group, \rev{along with the variance parameter $ \sigma_{u_i}^2 $}.  This extension to the literature of fitting several different correlation length parameter values strikes a balance between the stability of the maximum likelihood process (which can become very challenging were we to include 31 separate correlation lengths for each of the 31 input components) and the overall complexity of the residual process.  \rev{It should be noted that, whilst the maximum likelihood approach used here uses distributions to make an assessment of the correlation length parameters, this is only as a tool for calculating adequate Bayes linear emulators which satisfy diagnostics.}  At wave 8 we introduced the Dataset $ B $ outputs by first using linear model emulators for the new outputs only, and then using emulators with residual correlation processes for all outputs.  In waves 12 and 13 we incorporated emulators with residual processes for the Dataset $ C $ outputs.  

The number of design points per wave was largely kept constant at 2000.  2000 was deemed a suitable number of runs per wave as it meant that the matrix calculations involved in the emulator were reasonable, whilst allowing adequate coverage of the non-implausible input space with simulator runs.  At waves 11 and 13, 3500 design runs were used to build more accurate emulators.  

In terms of design, a maximin Latin hypercube \citep{CTM, BPDF} was deemed sufficient for our needs as we required a simple and efficient space-filling design.  The speed of the simulator meant that more structured and tactical designs were unnecessary for our requirements.  At wave 1 we constructed a Latin hypercube of size 2000 to build emulators for each of the outputs.  At waves 2-7 we first built a large maximin Latin hypercube design containing a large number of points over the smallest hyper-rectangle enclosing the non-implausible set.  We then used all previous wave emulators and implausibility measures to evaluate the implausibility of all the proposed points in the design \citep{GFBUA}.  Any points that did not satisfy the implausibility cut-offs were discarded from further analysis.  If a single Latin hypercube was not sufficient to generate enough design points, multiple Latin hypercubes were taken in turn and the remaining points in each were taken to be the design.  From wave 8 onwards an alternative sampling scheme was necessary to generate approximately uniform points from the non-implausible sets, since generating points using Latin hypercubes became infeasible due to the size of the non-implausible space.  There are several alternative ways presented in the literature to approximately sample uniformly distributed points over the non-implausible space \rev{\citep{BHMCIDM, HMHIV, EUDMWCE, SSHMSS}}.  We used a simple Metropolis-Hastings MCMC algorithm~\citep{brooks2011handbook}, which provided adequate coverage of the non-implausible space.

At each wave we performed diagnostics on the mean function linear model, the emulator and the implausibility criteria using 200 points in the non-implausible space.  The diagnostic test for implausibility which we used was the one described in \citep{BUCSBM}.  It compares the data implausibility cut-off criteria $ I^{data}_M(x) $, that is the implausibility evaluated at a known diagnostic run, against the chosen implausibility cut-off criteria for the emulator outputs.  

Many waves were necessary to complete this history matching procedure due to the complex structure of the Arabidopsis model.  We see from the first few waves that linear model emulators are sufficient for learning a great deal about the input parameter space, but that including full emulators with correlated residuals at later waves can be useful.  \rev{In addition, emulators have greatly increased the efficiency of our analysis over simply running the model.  To make a comparison, the model itself takes 30 seconds to run on a standard laptop computer (not very slow, but slow enough to cause problems) in comparison to the tens of millions of runs per second for the linear model emulators and tens of thousands of runs per second for the more complex later-wave emulators.} At the end of the procedure we obtained many runs satisfying each of the datasets $ A, B $ and $ C $.  We now go on to describe the results of the parameter search using various graphical representation techniques and discuss their biological implications.


\subsection{Visual Representation of History Matching Results}

A major aim of this work is to evaluate \rev{how acceptable parameter combinations to a model can be assessed as a result of experimental measurements}.

\begin{figure*}
\center
\includegraphics[width=12cm]{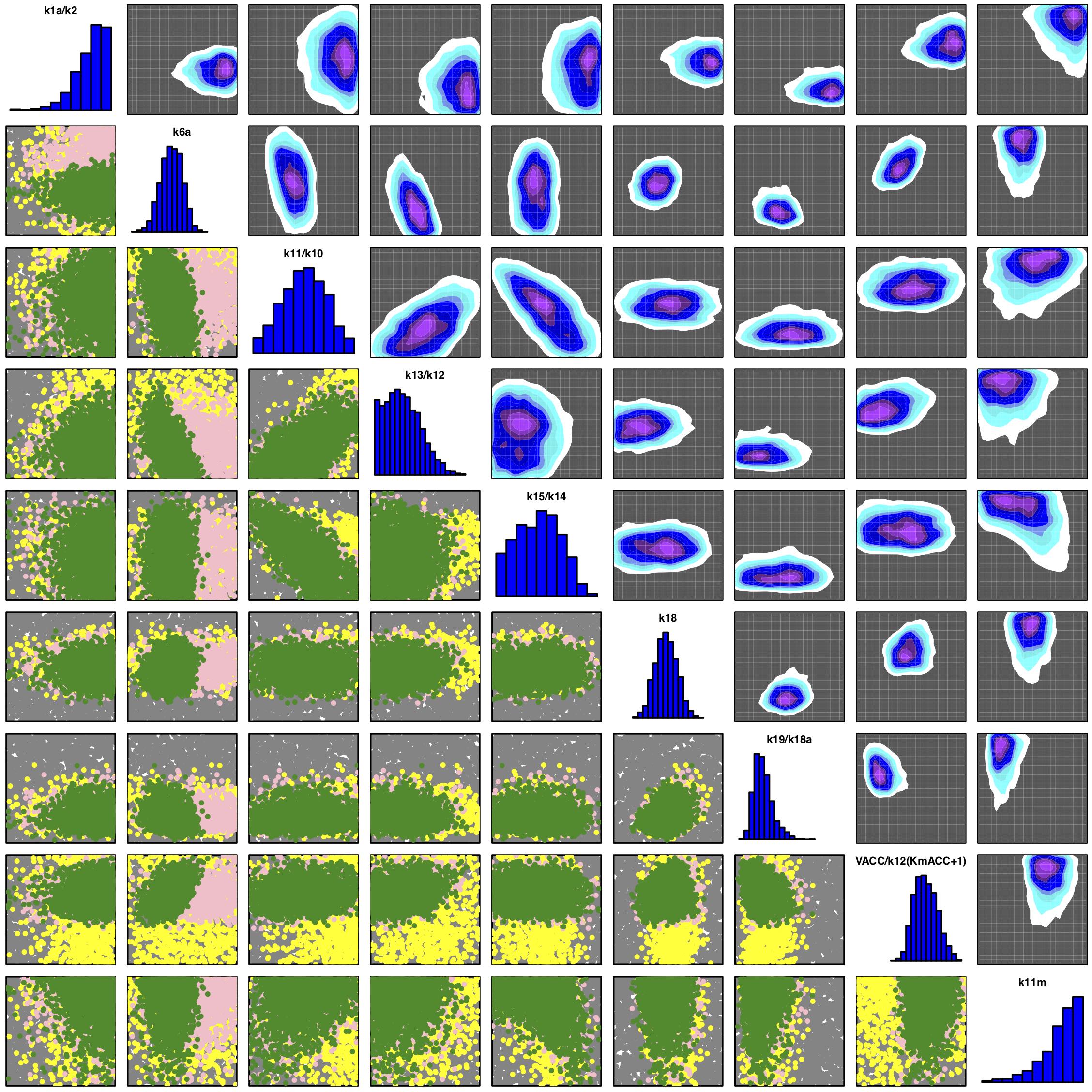}
\caption[2D Input Plot of Simulations]{Below diagonal: A pairs plot for a subset of the inputs $ x $.  Inputs to wave 1 runs are given by grey points.  Inputs to runs of the simulator with acceptable matches to the observed data in Datasets $ A $, $ B $ and $ C $ are given as yellow, pink and green points respectively.  Above diagonal: 2-dimensional optical density plots of inputs to runs with acceptable matches to all of the observed data for the same subset of the inputs.  The orientation of these plots has been flipped to be consistent with the plots below the diagonal.  Along diagonal: 1-dimensional optical density plots.  \label{IP2dsim}}
\end{figure*}

Figure \ref{IP2dsim} shows, below the diagonal, a ``pairs'' plot for a subset of the inputs $ x $.  A ``pairs'' plot shows the location of various points in the 31-dimensional input space projected down into 2-dimensional spaces corresponding to two of the inputs.  For example, the bottom left panel shows the points projected onto the $k_{1a}/k_2$ vs $k_{11m}$ plane.  Inputs to wave 1 runs are given by grey points.  Inputs to runs of the simulator with acceptable matches to the observed data in Datasets $ A $, $ B $ and $ C $ are given as yellow, pink and green points respectively.  Above the diagonal are shown 2-dimensional optical depth plots of inputs to runs with acceptable matches to all of the observed data for the same subset of the inputs.  Optical depth plots show the depth or thickness of the non-implausible space in the remaining 29 dimensions not shown in the 2d projection \citep{GFBUA,BUCSBM}.  \rev{More formally, suppose we partition input $ x $ as $ x = (x', x'') $, where $ x' $ is the two-dimensional vector representing the parameters we wish to project onto, and $ x'' $ represents the remaining 29  parameters, then the optical depth plot is given by:
\begin{equation}
\dist(x') \propto V(x \in \setx_C | x' \,\, \mbox{fixed} )
\end{equation}  
where $  V $ here represents volume in the remaining 29 dimensions.}  The orientation of these plots has been flipped to be consistent with the plots below the diagonal.  Along the diagonal are shown 1-dimensional optical depth plots. 

Figure \ref{IP2dsim} provides much insight into the structure of the model and the constraints placed upon the input rate parameters by the data.  Some of the inputs, such as $ k_{6a} $, $ k_{18} $, $ k_{19}/k_{18a} $ and $ V_{ACC}/k_{12}(Km_{ACC} + 1) $ are constrained even in terms of 1-dimensional range.  Some inputs only appear constrained when considered in combination with other inputs, for example $ k_{11}/k_{10} $ and $ k_{13}/k_{12} $ exhibit a positive correlation.  This is reasonable, since an increase in $ k_{11} $, the rate constant for converting the activated form of ethylene receptor into its inactivated form, can be compensated by an increase in $ k_{13} $, the rate constant for removing ethylene, since ethylene promotes the conversion of the activated form of ethylene receptor into its inactivated form.  More complex constraints involving three or more inputs are more difficult to visualise.  Below the diagonal, the pairs plot gives insight into which input parameters were learnt about by which set of outputs.  For example, the parameter $ V_{ACC}/k_{12}(Km_{ACC} + 1) $ is largely learnt about by Dataset $ B $, as is clear from the difference between the area of the yellow points and pink points in plots involving this input.  This is not surprising, since this term corresponds to the feeding and biosynthesis ($ k_{12} $) of ethylene, which we would expect to be learnt from the feeding ethylene experiments.  We can see that input combinations with large values of $ k_{6a} $ are classed as implausible, thus constraining this input to be relatively low. 
 
%
%
\begin{figure*}
\center
\includegraphics[width=12cm]{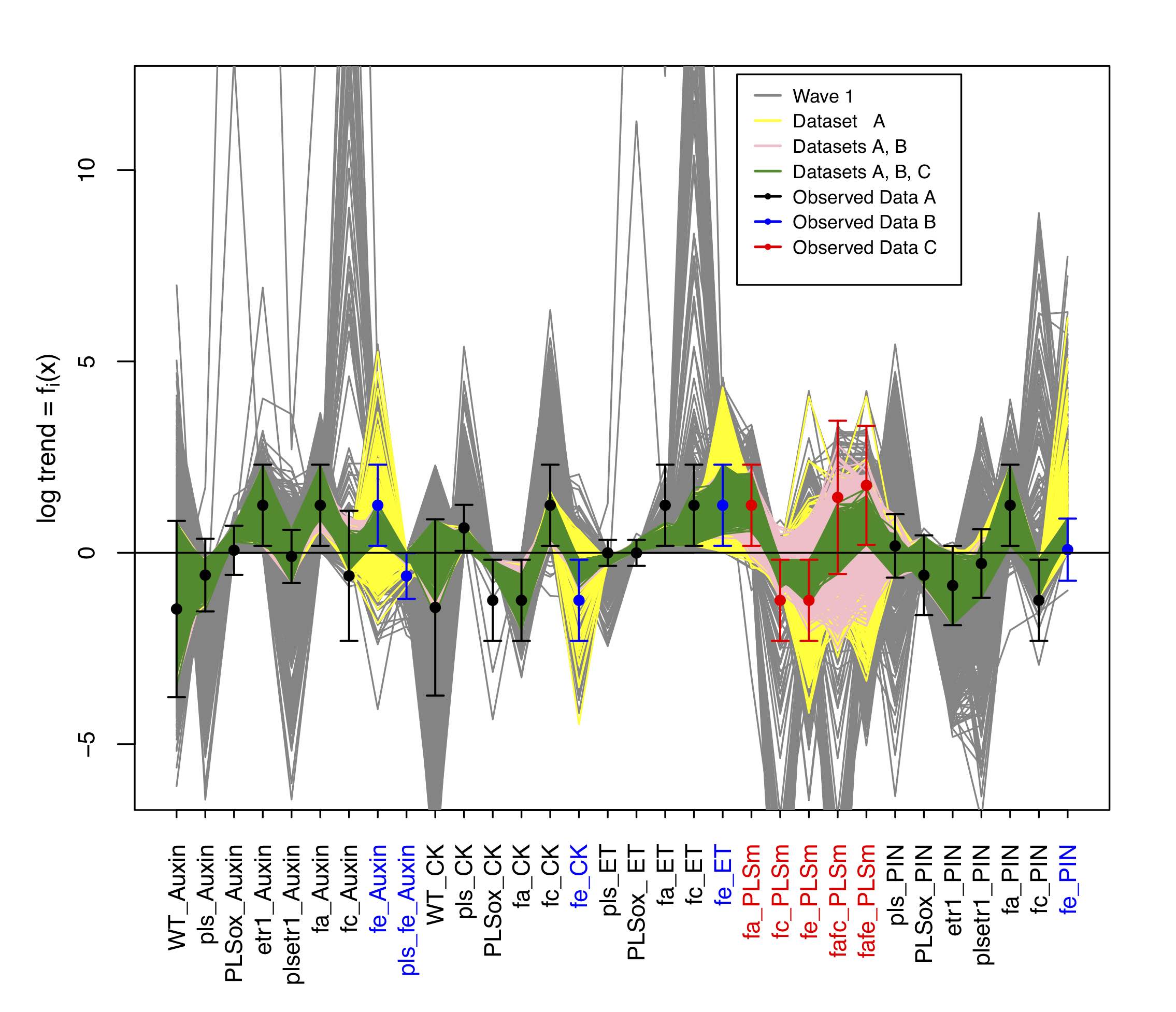}
\caption[1D Output Plot of Simulations]{Output runs $ f_i(x) $ for all 32 outputs considered.  Wave 1 runs are given as grey lines.  Simulator non-implausible runs after history matching Datasets $ A $, $ B $ and $ C $ are given as yellow, pink and green lines respectively.  The targets for the history match, as given by the intervals $ z_i \pm 3 \sigma_{c_i} $ and the ranges in Table \ref{OR}, are shown as vertical error bars.  Black error bars represent Dataset $ A $ outputs, blue error bars represent Dataset $ B $ outputs and red error bars represent Dataset $ C $ outputs.  The horizontal black line at zero represents zero trend.  \label{OP1dsim}}  
\end{figure*}

Figure \ref{OP1dsim} shows the output runs $ f_i(x) $ corresponding to the input combinations shown in Figure \ref{IP2dsim} for all 32 outputs considered.  The colour scheme is directly consistent with Figure \ref{IP2dsim}, with  wave 1 runs given as grey lines, and simulator non-implausible runs after history matching Datasets $ A $, $ B $ and $ C $ given as yellow, pink and green lines respectively.  Runs which pass within the error bar of a particular output $ i $ satisfy the constraint of being within $ z_i \pm 3 \sigma_{c_i} $, thus being in alignment with the results of the corresponding experimental observation, given our beliefs about model discrepancy and measurement error.  Black error bars represent Dataset $ A $ outputs, blue error bars represent Dataset $ B $ outputs and red error bars represent Dataset $ C $ outputs.  

Figure \ref{OP1dsim} gives much insight into joint constraints on possible model output values that are in alignment with all of the observed data (and so would pass through all of the error bars).  Some model outputs have been constrained much more than the range of their error bars, for example, $ PLSox\_Auxin $ is constrained to the upper half of its error bar while $ f_c\_CK $ is constrained to take smaller values.  It is interesting that many of the yellow runs already go through the error bars of some of the outputs in Datasets $ B $ and $ C $, for example $ pls\_f_e\_Auxin $ and $ f_a\_PLSm $.  This indicates that the additional experimental observations corresponding to such outputs did not help to further constrain the input space.  


\begin{figure*}
\center
\includegraphics[width=12cm]{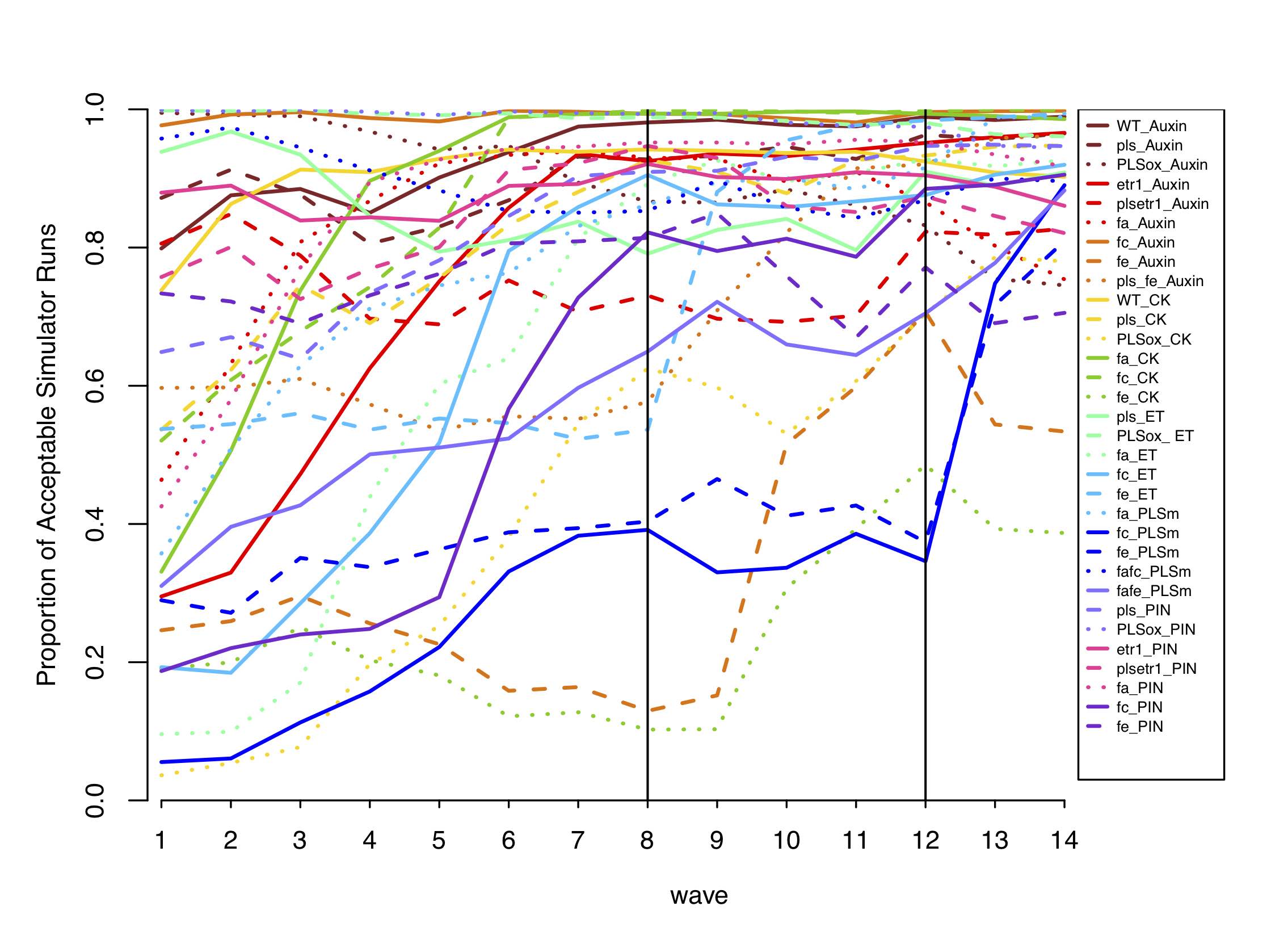}
\caption[Proportion of non-implausible runs by output]{The proportion of simulator runs at each wave which pass through the error bar of each output. 
The two vertical black lines represent the waves where datasets B and C were incorporated.  \label{PNIO} }
\end{figure*}

Figure \ref{PNIO} presents the proportion of simulator runs at each wave which pass through the error bar of each output.  Lower numbers for a particular output at a particular wave indicate that the output could be informative for learning more about the input parameter space.  The two vertical black lines represent the waves where datasets B and C were incorporated. Some outputs, for example $ PLSox\_ET $ and $ PLSox\_PIN $, had a high proportion (close to 1) of runs passing through their error bars at wave 1, in accordance with Figure \ref{IW1R}.  These outputs were not very informative for the history matching process.  Some outputs, for example, $ etr1\_Auxin $ and $ f_c\_PLSm $, had a low proportion (0.29 and 0.08 respectively) of runs passing through their error bars at wave 1, but a high proportion (over 0.8) after 13 waves of history matching.  Space that would be classed as implausible by these simulator outputs became classed as implausible by the emulators during the waves of history matching.  Some outputs, for example $ f_e\_Auxin $ and $ f_e\_CK $, had relatively low proportions (less than 0.6) of runs passing through their error bars even at the end of the history matching procedure.  This is indication that these outputs may have been difficult to emulate throughout.  


As expected, we notice that the outputs in Datasets $ B $ and $ C $ start to have higher numbers of runs passing through their error bars once those outputs have been history matched to observations.  Interestingly, as can also be detected in Figure \ref{OP1dsim}, some of the outputs in Datasets $ B $ and $ C $, for example $ f_af_e\_PLSm $, get a surprisingly increased proportion (from 0.32 to 0.63) of runs passing through their error bars even before the output is incorporated into the history match.  This is an indication that information from this output has already been learnt from observing some combination of the previously included outputs.  \rev{There are a few components, most notably $ PLSox\_Auxin $, which had a high proportion of runs passing through their error bars before wave 1, but a much smaller proportion by the end.  This is possible due to the joint constraints between the output components which involves non-implausible runs for this output component being classed as implausible by the constraints related to other output components.  In addition, such behaviour is much less surprising if a particular output component was not included in the history match at early waves.}


Although the overall proportion of space cut out is a very useful measure of the dependence of the model input parameter space on observed measurements, one may be interested in the degree to which specific parameters of particular interest have been constrained due to the observations.  Sample variances of particular inputs in the non-implausible sets are a very informative and appropriate measure for this purpose as they take account of 
the density of the non-implausible space projected down onto the input dimensions of specific interest. Such measures are simple to calculate, and in many cases sufficient for our purposes.  However, if we wanted to perform a full Bayesian analysis, we could appropriately re-weight the non-implausible points and recalculate these sample variances to obtain estimates of posterior (marginal) variances, provided we were confident enough to make all the additional assumptions that a full Bayesian analysis requires, as outlined in Section \ref{BHMMCMC}.

\begin{figure*}
\center
\includegraphics[width=12cm]{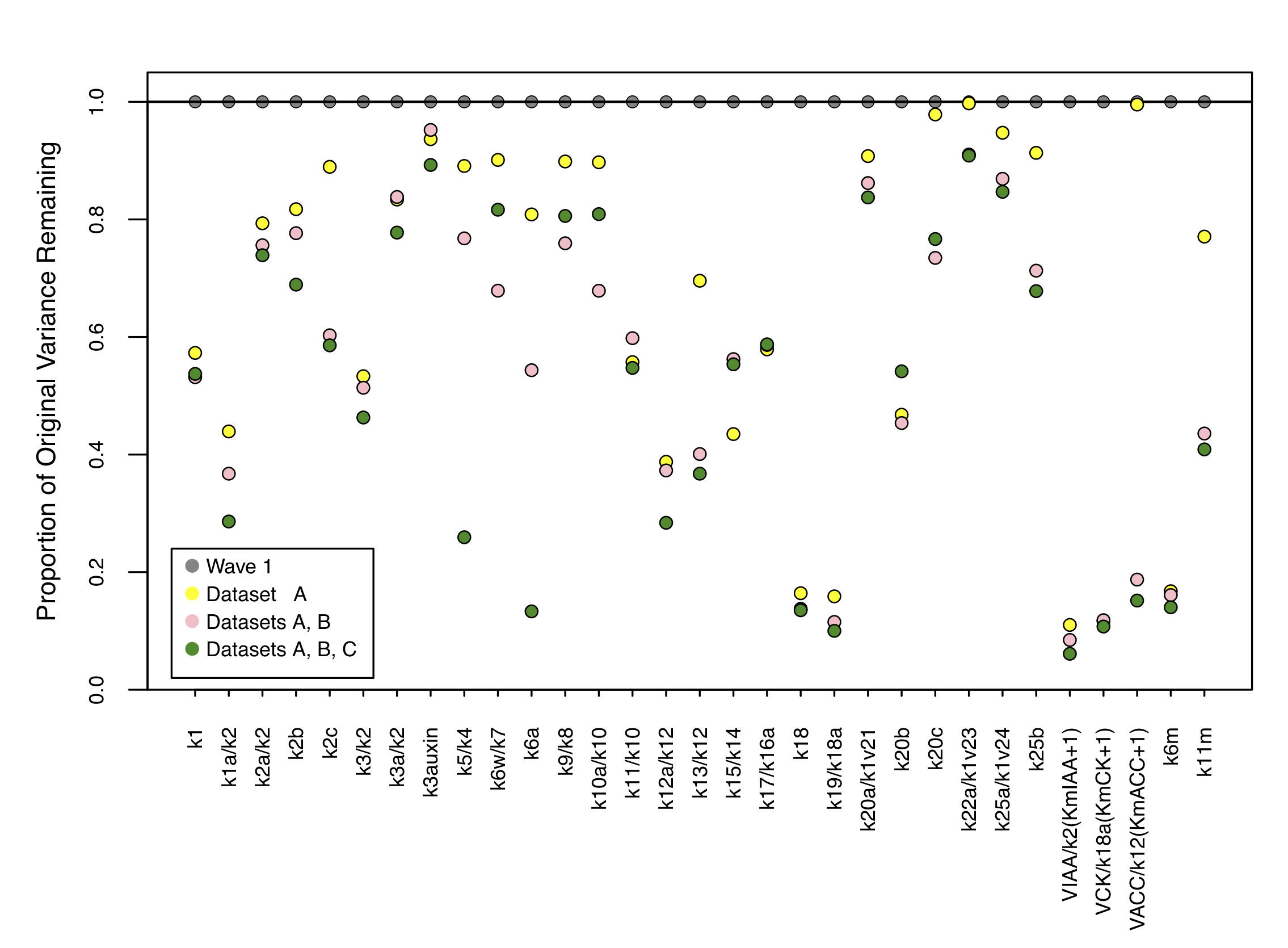}
\caption[1D Variance Plot of the Inputs]{Sample variances (as a proportion of the original wave 1 sample variance) for each input of a sample of 2000 points with acceptable matches to the observed data in Datasets $ A $, $ B $ and $ C $, given by yellow, pink and green points respectively. The difference between the grey (wave 1) and yellow points shows the proportion of sample variance resolved by the Dataset A outputs.  The differences between the yellow and pink, and pink and green points show the amount of additional space resolved (as a proportion of the original sample variance) by the Datasets B and C outputs respectively.  \label{VR1dsim}}
\end{figure*}

In Figure \ref{VR1dsim}, sample variances (as a proportion of the original wave 1 sample variance) for each input of a sample of 2000 points with acceptable matches to the observed data in Datasets $ A $, $ B $ and $ C $ are given by yellow, pink and green points respectively.  Again, there is much insight to be gained from such a plot. We can see that different input ratios have been learnt about to different degrees by the observations of outputs in Datasets $ A $, $ B $ and $ C $.  Some inputs are resolved well by Dataset $ A $ but then not really any further once Datasets $ B $ and $ C $ are additionally introduced.  For example, $ k_1 $, representing inhibition of auxin transport by the ethylene downstream, $ X $, is reduced by 0.43 by Dataset $ A $, and then by less than 0.1 after both $ B $ and $ C $ have been additionally measured. This implies that experiments related to feeding ACC and measuring the PLS gene expression play a limited role in determining the parameter about inhibition of auxin transport by ethylene downstream.  Some inputs are resolved slightly by Datasets $ A $ and $ B $, and then substantially by Dataset $ C $.  For example, $ k_5/k_4 $, which governs the rate of conversion of auxin receptor from its active form $ Ra $ to its inactive form $ Ra^\ast $ and vice-versa, is reduced by less than 0.25 by Datasets $ A $ and $ B $, and then by more than an additional 0.5 once Dataset $ C $ is measured.  By analysing the model equations we see that $ [Ra] $ and $ [Ra^\ast] $ feature prominently in the $ [PLSm] $ equation, which is the output being measured in Dataset C. This indicates that measuring the $ PLS $ gene expression is important for determining the parameter relating to activation and inactivation of auxin receptor. Some inputs, for example $ k_{6a} $, are learnt partially about by each dataset in turn, with overall high resolution.  Some inputs have very little variance resolution at all.  For example, $ k_{22a}/k_{1_v23} $, representing $ PIN1m $ translation to produce $ PIN1pi $, has an approximate resolution of 0.1. Some information contained in Figure \ref{VR1dsim} may be quite intuitive, 
for example the fact that most of the variance resolution of $ V_{ACC}/k_{12}(Km_{ACC}+1) $, the input corresponding to the feeding of ethylene, is obtained after measuring Dataset $ B $.  Checking that our results coincide with this intuitive biological knowledge is an important diagnostic step, and provides evidence that our method has analysed the parameter space appropriately.  Other information contained in Figure \ref{VR1dsim} is less intuitive and offers insight into the complex structure of the Arabidopsis model.

\begin{figure*}
\center
\includegraphics[width=12cm]{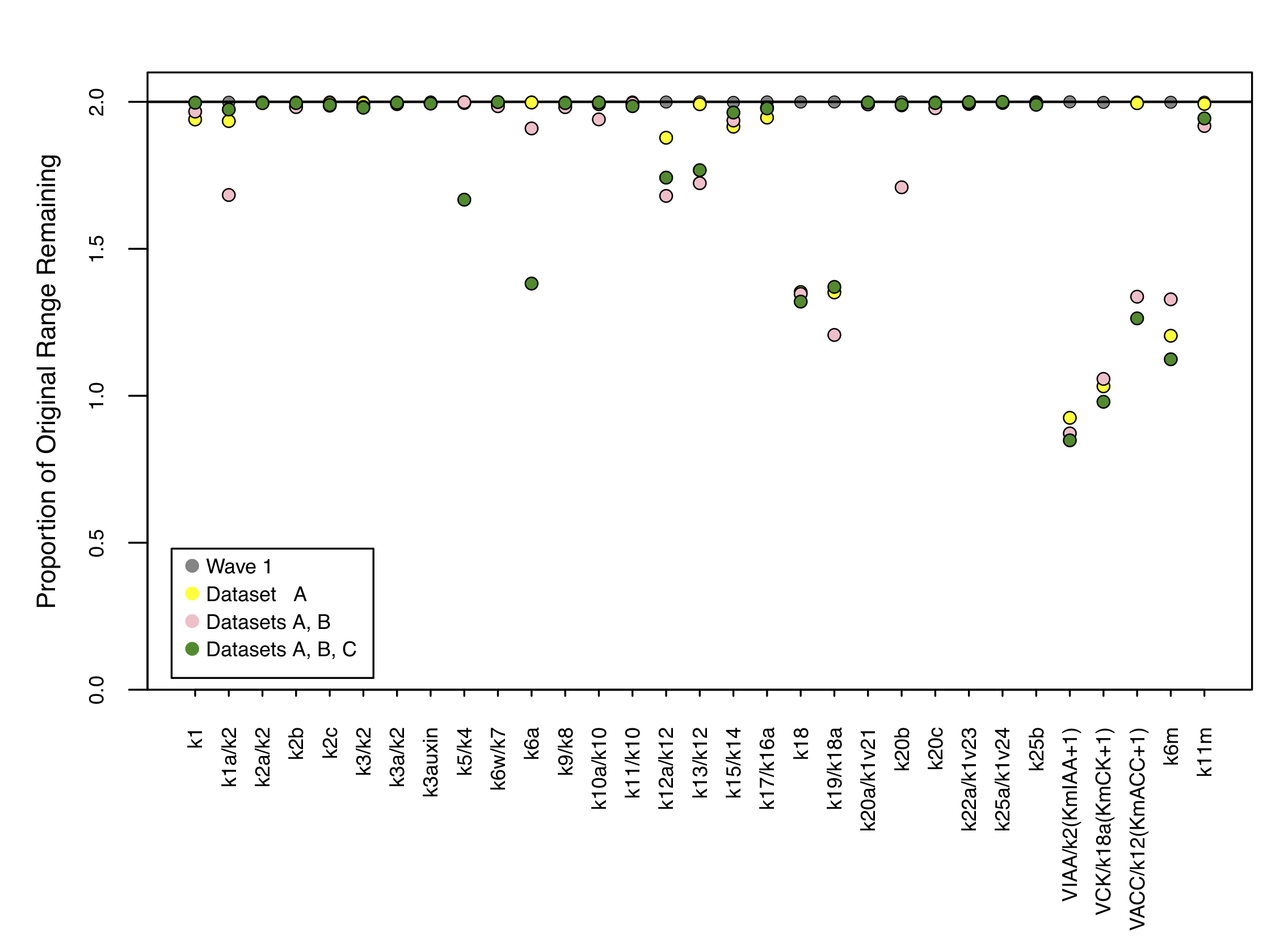}
\caption[1D Range Plot of the Inputs]{Sample ranges for each input of the runs used to build the wave 1 emulator are shown as grey points.  Ranges for each input of a sample of 2000 points with acceptable matches to the observed data in Datasets $ A $, $ B $ and $ C $ are given by yellow, pink and green points respectively.  \label{RR1dsim} }
\end{figure*}

An analogous measure to space cut out in lower dimensions is range, area or volume reduction of the non-implausible space projected down onto the relevant input dimensions.  These measures are far less informative than variance measures as they are very sensitive to extreme values, and it is not uncommon for the initial range of an input to be non-implausible in high dimensions.  To get an idea of this, we compare Figure \ref{VR1dsim} to Figure \ref{RR1dsim}, which presents ranges for each input of a sample of runs used to build the wave 1 emulator as grey points, and ranges for each input of a sample of 2000 points with acceptable matches to the observed data in Datasets $ A $, $ B $ and $ C $ as yellow, pink and green points respectively.  We can see that certain inputs, for example $ k_{19}/k_{18a} $, $ k_{6m} $ and in particular the feeding inputs $ \frac{V_{IAA}[IAA]}{Km_{IAA} + [IAA]} $, $ \frac{V_{CK}[cytokinin]}{Km_{CK} + [cytokinin]} $ and $ \frac{V_{ACC}[ACC]}{Km_{ACC} + [ACC]} $, have their ranges significantly reduced.  Many of the other inputs don't have their sample ranges reduced much at all.  This does not necessarily mean that we don't learn about these inputs, just that for any specified value of one of these inputs there exists some combination of the remaining 30 inputs which can compensate, hence leading to a model output with an acceptable match to the observed data. 

Simple measures involving the analysis of variance reduction or resolution can also be used to quickly describe joint constraints that alert us to strong relationships between inputs.  Suppose we treat the vector of inputs as a multi-dimensional random variable $ \rv^u $ uniformly distributed over a non-implausible region $ \setx_u $, that is:  
$$ f_{W^u}(w^u) \propto \left\{ \begin{array}{cc} 1, & w^u \in \setx_u \\ 0, & w^u \not\in \setx_u \\ \end{array} \right. $$  
Note that the uniform distribution is chosen here as we wish to treat all parts of the non-implausible set equally, as we may currently doubt the existence of a ``true" best input $ \bestx $, and hence not want to perform a posterior re-weighting of the region $\setx$.
Given $ f_{W^u}(w^u) $, we can calculate $ \variance[W^u] $.  Let us define the marginal variance for inputs $ J = j_1,..., j_m $ of random variable $ \rv^u $ corresponding to non-implausible space $ \setx_u $ as $ \variance[\rv_J^u] $.  We introduce the variance resolution measure for inputs $ J $ between non-implausible spaces $ \setx_u $ and $ \setx_v $ to be:
\begin{equation}
R_{uv}(\setx_J) = 1 - \frac{\det(\variance[\rv^v_J])}{\det(\variance[\rv_J^u])} 
\end{equation}
We choose this measure as it relates to the product of the eigenvalues of the variance matrix and hence to a density-weighted volume of the projected non-implausible space, which is relevant for what we are interested in. We do not have exact distributions for $ f_{W^u}(w^u) $ owing to not having an exact specification for $ \setx_u $.  We therefore estimate $ \variance[\rv^u] $ corresponding to $ \setx_u $ as $ \variance[\setx^s_u] $, where $ \setx_u^s $ is an (approximately) uniform sample of points from the non-implausible set $ \setx_u $.

\begin{figure*}
\center
\includegraphics[width=12cm, height=13.3cm]{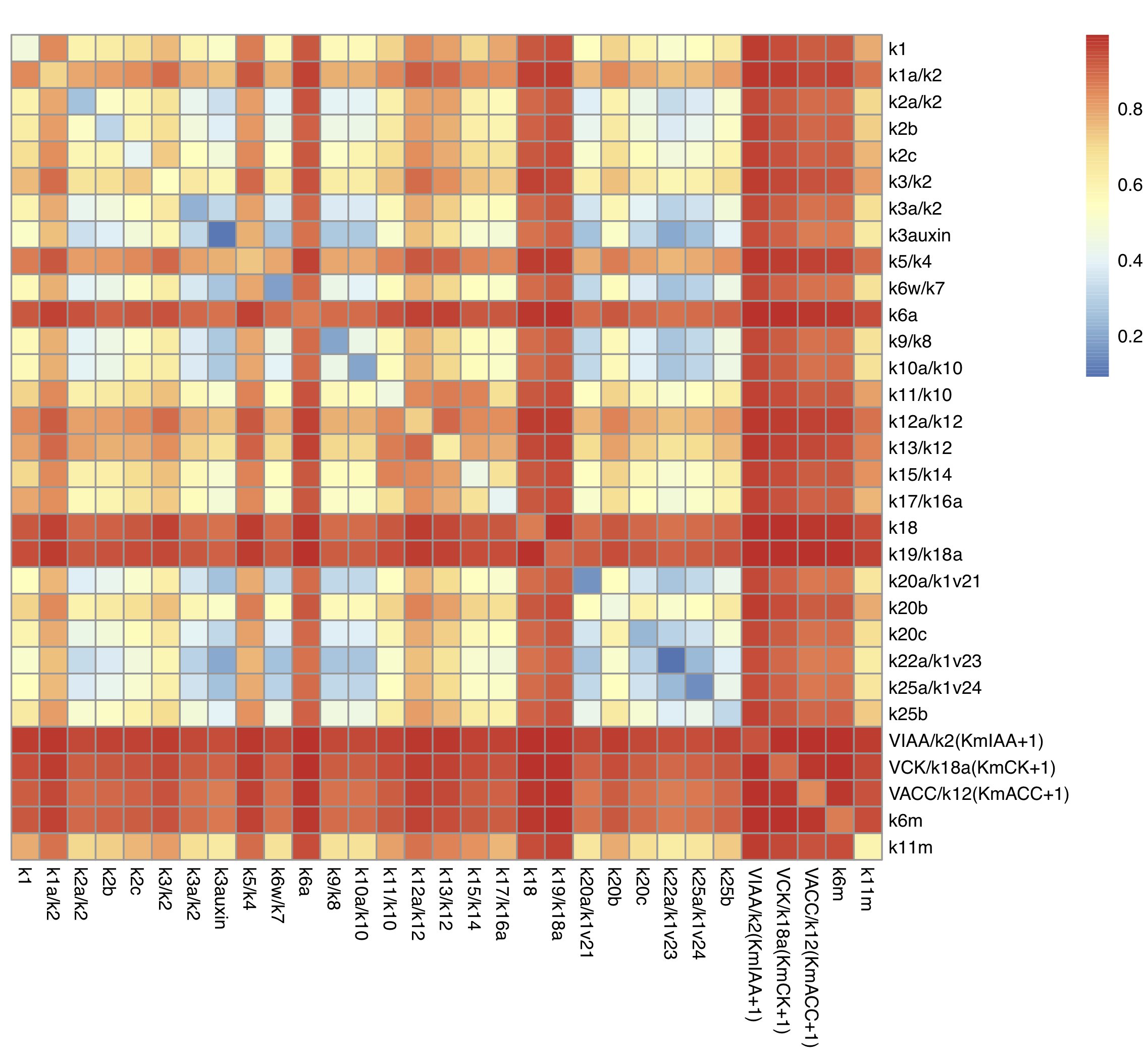}
\caption[2D Variance Resolution Plot]{Sample variance resolutions $ R_{0C}(\setx_{j_1,j_2}^s) $  between initial and final non-implausible spaces for each pair of inputs $ j_1, j_2 $, represented by colour.  Individual input variances, corresponding to the difference between the grey points and green points in Figure \ref{VR1dsim}, are represented along the diagonal.  Red indicates high resolution whereas blue represents low resolution.  For example, a large amount of variance has been jointly resolved by $ k_{6a} $ and $ k_5/k_4 $, as indicated by the red square at the intersection of the corresponding row and column. \label{VR2dsim} }
\end{figure*}

Figure \ref{VR2dsim} shows sample variance resolutions $ R_{0C}(\setx_{j_1,j_2}^s) $ between initial (0) and final (C) non-implausible spaces for each pair of inputs $ J =  j_1,j_2 $, represented by colour, with red indicating high resolution and blue representing low resolution.  Individual input variance resolutions, namely the difference between the initial grey and final green points in Figure \ref{VR1dsim}, are represented along the diagonal.  Note that an individual input resolution will never be greater than the joint variance resolution of that input with another one.  We can see that learning jointly about $ k_{1a}/k_2 $ and $ k_{18} $, namely those rate parameters representing auxin transport and biosynthesis, and regulation of cytokinin biosynthesis by auxin, is seemingly more informative than learning about either of the two parameters separately, in terms of variance resolution. As a converse example, we can see that little is learnt jointly between $ k_{3auxin} $ and $ k_{22a}/k_{1_v23} $.

\begin{figure*}
\center
 \includegraphics[width=12cm, height=12cm]{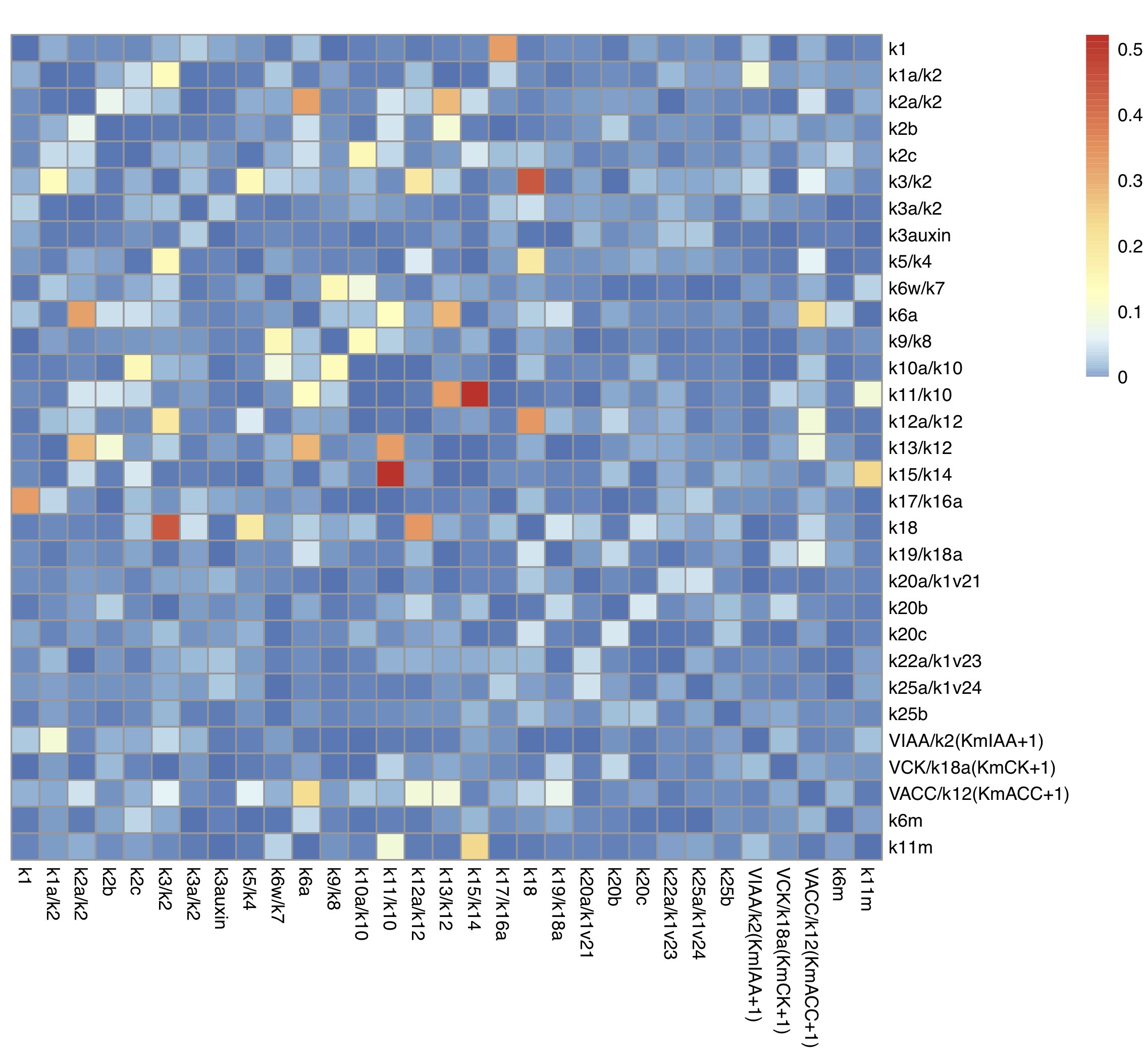}
\caption[2D Variance Difference Plot]{Standardised differences between expected sample variance assuming independence between the inputs and the actual variance in the final non-implausible space for each pair of inputs, represented by colour.  The diagonal elements are zero.  Red represents a larger difference and blue represents a smaller difference.  For example, there is a strong joint constraint between $ k_{11}/k_{10} $ and $ k_{15}/k_{14} $, and little joint constraint between $ k_{18} $ and $ k_{1a}/k_2 $.
 \label{VD2dsim}}
\end{figure*}

Although Figure \ref{VR2dsim} is informative, we really wish to determine the cases where the joint constraint on two input parameters is more severe than we would expect if we just assumed they were independently constrained. 
Assuming independence, the determinant of the sample variance matrix for inputs $ j_1, j_2 $ is the product of the sample variance for each input, that is:
\begin{equation}
\det(\variance[\setx^s_{j_1, j_2}]) = \variance[\setx^s_{j_1}]\variance[\setx^s_{j_2}] 
\end{equation}
The standardised difference between the determinant assuming independent inputs and observed determinant is the squared correlation function:
\begin{equation} 
\frac{\variance[\setx^s_{j_1}]\variance[\setx^s_{j_2}] - \det(\variance[\setx^s_{j_1,j_2}])}{\variance[\setx^s_{j_1}]\variance[\setx^s_{j_2}]} = \frac{(\covariance[\setx^s_{j_1}, \setx^s_{j_2}])^2}{\variance[\setx^s_{j_1}]\variance[\setx^s_{j_2}]} 
\end{equation}
This is informative for the dependence, and hence level of constraint, between that pair of inputs in the final non-implausible set, as it alerts us to cases
where there has been far more variance resolved jointly than that expected were the inputs just constrained independently.  We therefore present these differences between each pair of inputs, represented by colour, in Figure \ref{VD2dsim}.  Red represents a larger difference and blue represents a smaller difference.  The diagonal elements are necessarily zero.  It would appear that there are stronger joint constraints between lots of pairs of inputs, most notably $ k_{11}/k_{10} $ and $ k_{15}/k_{14} $, and $ k_3/k_2 $ and $ k_{18} $.  The first of these pairs, involving the CTR1 protein and ethylene receptor, has the most joint structure of any pair, with a squared sample correlation of 0.46. Since both the CTR1 protein and ethylene receptor take actions in the ethylene signalling module, they relay ethylene signalling.  The parameters controlling this relay can be highly inter-dependent.  Therefore, a change in one of these parameters can be compensated by a change in another. Interestingly, Figure \ref{VD2dsim} would indicate that there is little joint structure between $ k_{18} $ and $ k_{1a}/k_2 $, with a squared sample correlation of less than 0.05, indicating that the combined variance resolution between $ k_{1a}/k_2 $ and $ k_{18} $ presented in Figure \ref{VR2dsim} was not much larger than the independent product of the resolution of each of the individual inputs.
Figure \ref{VD2dsim} can suggest interesting pairs of inputs to analyse in more detail, for example by examining their corresponding pairs plots, as 
shown in Figure \ref{IP2dsim}.

\begin{figure*}
\center
\includegraphics[width=12cm, height=12cm]{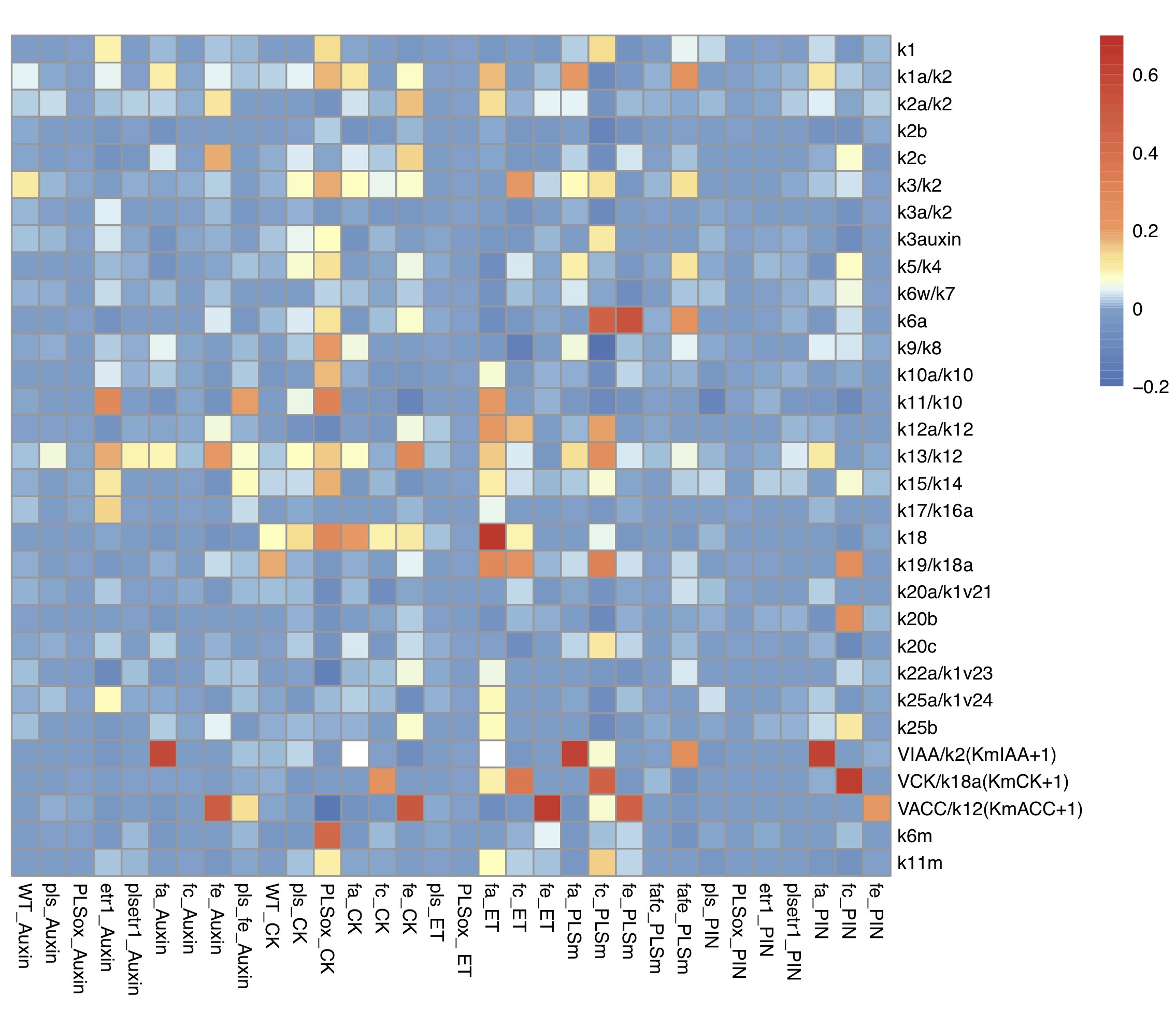}
\caption[Input-Output Variance Resolution Plot]{Estimates of how much each output informed each input, represented by colour.  These were calculated as the standardised difference between the sample variance of the input for all wave 1 runs and those wave 1 runs going through the output error bar.  Red indicates higher values of this estimated quantity and blue represents lower values.\label{IOVRP}}
\end{figure*}

Figure \ref{IOVRP} provides a visualisation of how much each single output informed each input, represented by colour.  These were calculated as the standardised difference between the sample variance of the input for all wave 1 runs and those wave 1 runs going through the output error bar.  This quantity estimates the sample variance resolution for each input $ i $ were we to history match using only output $ j $.  Red indicates higher values of this estimated quantity and blue represents lower values.  Figure \ref{IOVRP} is very informative.  We can see that some of the outputs, for example $ PLSox\_Auxin $ and $ f_c\_Auxin $, are not directly informative about many of the inputs when considered on their own (however this does not preclude the possibility that they could still be informative about the input parameters when measured in conjunction 
with other outputs).
This is in alignment with Figure \ref{PNIO}. Conversely, some outputs are very informative about a few specific inputs, for example $ f_a\_Auxin $ is particularly informative for $ k_{1a}/k_2 $, $ k_{13}/k_{12} $ and $ V_{IAA}/k_2(Km_{IAA}+1) $, with estimated sample variance resolutions of 0.14, 0.13 and 0.52 respectively.  It may be unsurprising that matching to the observation of auxin when feeding auxin is informative for learning about the rate parameters $ k_{1a}/k_2 $ or $ V_{IAA}/k_2(Km_{IAA}+1) $, which represent auxin transport to the cell and the quantity of auxin taken up by the plant respectively.  It is more interesting, however, that this experimental observation is also informative for learning about the parameter $ k_{13}/k_{12} $, representing the relationship between biosynthesis and decay of ethylene.  Other outputs, for example $ etr1\_Auxin $ and $ pls\_CK $, are slightly informative for a range of the inputs, but not very informative for any of them.  This indicates that these outputs are quite informative for learning about the rate parameters and their relationships with each other across the whole network.

Conversely, we can see from Figure \ref{IOVRP} that each input is informed about by a different variety of outputs.  Some inputs are learnt about by a large number of outputs, for example $ k_3/k_2 $ and $ k_{13}/k_{12} $ which are the decay rate parameters for the decay of auxin and ethylene respectively from the cell.  Interestingly, many of these outputs involved the measurement of cytokinin.  Some inputs, for example $ k_{2b} $ and $ k_{3a}/k_2 $, don't seem to be informed about by many outputs at all.  These results are in alignment with Figure \ref{VR1dsim} which shows the general change in variance for each input.  Other inputs are learnt about quite heavily by just a few outputs.  For example, $ k_{6m} $ which represents the additional $ PLS $ transcription rate in \emph{PLSox} relative to \emph{WT}, is learnt about heavily after measuring $ PLSox\_CK $, that is the measurement of cytokinin concentration under the mutant relative to that of wild type, with sample variance resolution 0.32.  We can see that such an analysis of which output measurements inform us about which input constraints can be insightful.  Some of the input-output relationships may be quite intuitive, whilst others inform us about links between the inputs and the outputs of which we were unaware before we started the history matching analysis.  Whilst Figure 12 is informative, it is limited to information about the relationship between one input and one output. Information about how single outputs inform us about complex interactions between inputs, or how multiple outputs may be telling us similar information about particular inputs, is not displayed.
	
\subsection{Gaining Insight Into Specific Learning Objectives From History Matching Results}

Insight into many specific aspects of the model of particular interest can be obtained from the results of a history match.  For example, some results in the literature suggest that output $ f_c\_Auxin $, that is measuring the ratio of Auxin concentration in wild type fed cytokinin relative to wild type with no feeding, is a down trend, whilst others suggest that it is an up trend \citep{CRAS}.  We therefore separate the final sample of acceptable runs into two groups to analyse whether further measurements of this output this would have an effect on our conclusions.

\begin{figure*}
\center
\includegraphics[width=12cm]{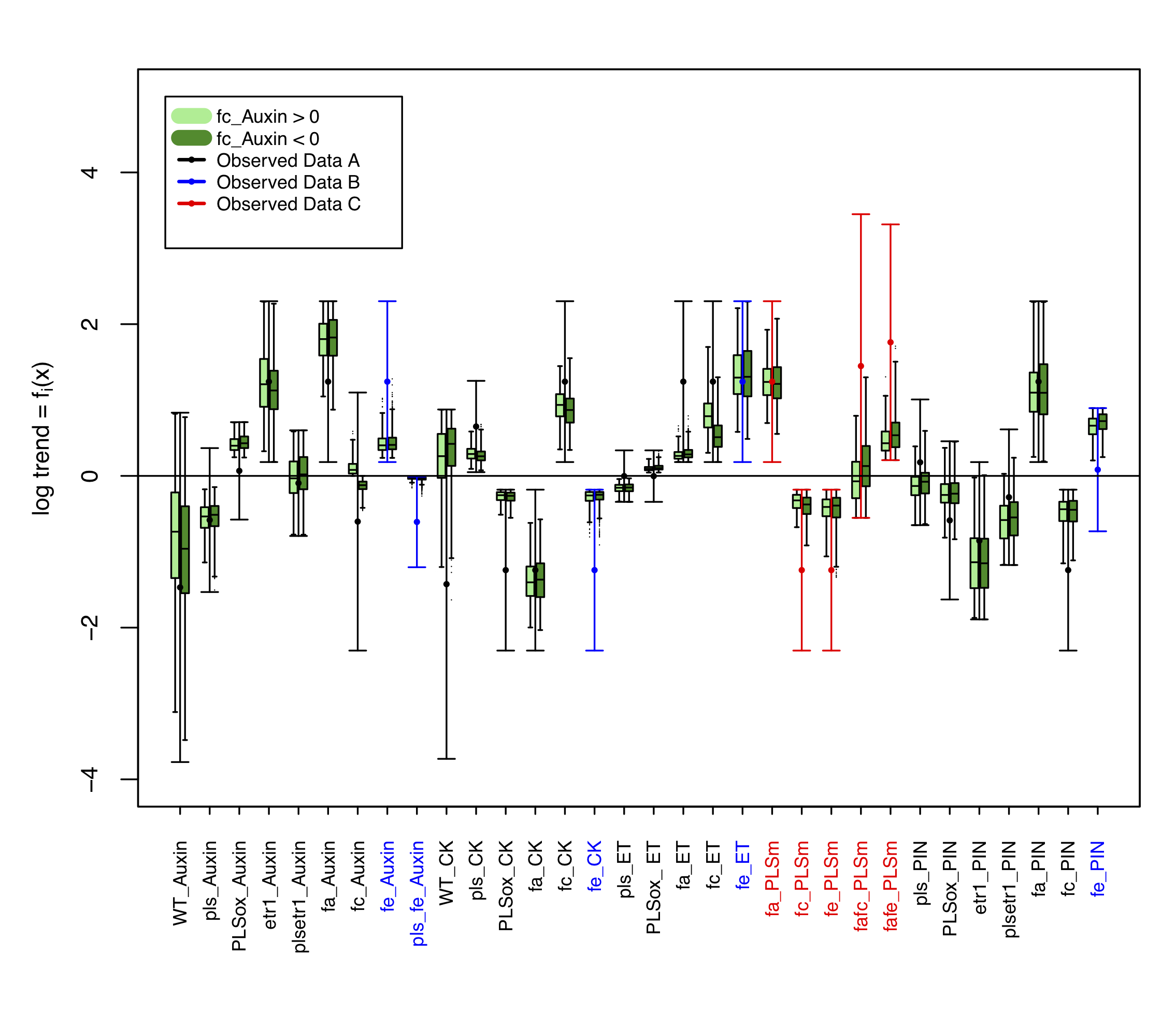}
\caption[1D Output Plot of Simulations]{Boxplots summarising the range of output values for simulator runs $ f_i(x) $ for all 32 outputs considered that satisfied all of the error bars.  The light green boxplots are for runs having positive output value for $ f_c\_Auxin $ and dark green boxplots are for runs having negative value for this output.  The targets for the history match, as given by the intervals $ z_i \pm 3 \sigma_{c_i} $ and the ranges in Table \ref{OR}, are shown as vertical error bars.  Black error bars represent Dataset $ A $ outputs, blue error bars represent Dataset $ B $ outputs and red error bars represent Dataset $ C $ outputs.  The horizontal black line at zero represents zero trend. \label{BoxplotsSim} }  
\end{figure*}

Figure \ref{BoxplotsSim} shows boxplots summarising the range of output values for simulator runs $ f_i(x) $ of all 32 outputs for the final sample of acceptable runs.  The light green boxplots are for runs having positive output value for $ f_c\_Auxin $ and dark green boxplots are for runs having negative value for this output.  Approximately 80\% of the sample runs in the final non-implausible input space have negative values for output $ f_c\_Auxin $ relative to approximately 45\% of the initial wave 1 runs.  This is a result of matching to other outputs since nearly all initial runs already went through the error bar for $ f_c\_Auxin $.  There are a few outputs, for example $ f_c\_ET $, which distinguish between runs with positive or negative values of $ f_c\_Auxin $.  However, in general, it would appear that most of the other outputs are relatively independent of $ f_c\_Auxin $.  Therefore, it could be worth taking more careful observations of experiment $ f_c\_Auxin $ in order to learn more about the effect of feeding cytokinin on auxin concentration.  Such observations may provide information about the model and physical system which is not being captured by the other experiments.    

\begin{figure*}
\center
\includegraphics[width=10cm]{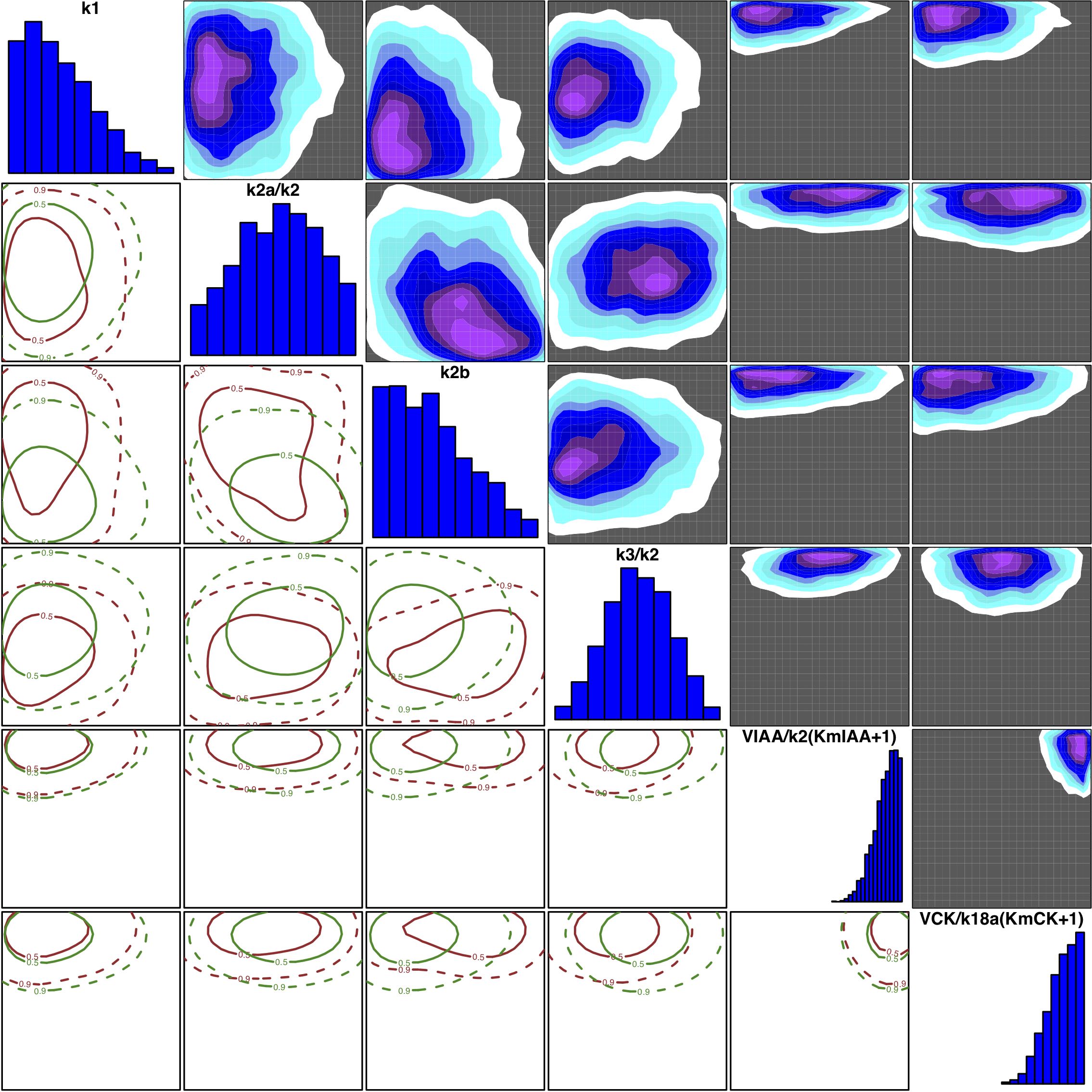}
\caption[2D Input Plot of Simulations]{Below diagonal: Contours showing the 0.5 and 0.9 highest density sets for an initial sample of wave 14 runs.  Brown contours indicate runs with positive value for output $ f_c\_Auxin $ and green runs have negative values for this output.  Above diagonal: 2-dimensional optical density plots of inputs to runs with acceptable matches to all of the observed data for the same subset of the inputs.  The orientation of these plots has been flipped to be consistent with the plots below the diagonal.  Along diagonal: 1-dimensional optical density plots. \label{IP2dsimdisw14}}
\end{figure*}

Figure \ref{IP2dsimdisw14} shows, below the diagonal, for each pair of a subset of inputs for the final simulator acceptable runs, the boundaries of the 0.5 and 0.9 highest density sets as solid and dashed contours respectively.  Brown contours indicate runs with positive value for output $ f_c\_Auxin $ and green runs have negative values for this output.  We can see that some inputs, for example $ k_{2b} $ and $ k_3/k_2 $ involving the effects of auxin and cytokinin concentrations on the rate of change of auxin concentration, tend to show a distinction between runs with positive and negative output values for $ f_c\_Auxin. $.  This suggests that further measurement of $ f_c\_Auxin $ would be informative for learning about these rate parameters.  Above the diagonal are the overall density plots for this subset of outputs for comparison.

Many other interesting features of the model could be analysed in a similar way.  In future work we will demonstrate how we can design future experiments using complex models, combined with history matching methodology, in order to choose the set of measurements to perform that will have the best chance of learning about specific scientific 
criteria of interest. 
	

\section{Discussion and Conclusions}

Understanding how hormones and genes interact to coordinate plant growth  is a major challenge in plant developmental biology. Auxin, cytokinin and ethylene are three important hormones that regulate many aspects of plant development.  The dynamics of this crosstalk are non-linear and unintuitive \citep{HCRD, CCACE}.  Experimental measurements are necessary in order to represent the general dynamics of such a system by formulating kinetic equations.  In particular, it is essential to establish how the associated model parameter space can be informed about by experimental observations, since understanding of the rate parameters is essential for a model to be informative for a physical system.  In this work, we have shown how comprehensive exploration and understanding of the input rate parameter space can be achieved from sets of experimental observations by applying sequential history matching techniques using Bayesian emulation. 


The rate parameter $ k_{6a} $ describes how the POLARIS transcriptional rate is regulated by ethylene \citep{MEAHCA}.  Increasing $ k_{6a} $ decreases the strength of this regulation.  Figures \ref{IP2dsim} and \ref{OP1dsim} suggest that the set of possible values of $ k_{6a} $ which satisfy all of the observed data is quite constrained, with large values and the smallest values in the initial range being classed as implausible. Noticeably, this parameter was primarily constrained by the inclusion of dataset C, which measured PLSm.

The parameter ratio $ k_{6w}/k_7 $ represents the transcription rate of the POLARIS gene function in wild type, and the parameter $ k_{6m} $ represents the additional POLARIS transcription when the POLARIS gene is overexpressed.  Figure \ref{VR1dsim} suggests that $ k_{6w}/k_7 $ is not highly constrained by the observed measurements, but that $ k_{6m} $ is highly constrained after history matching to the observations in Dataset $ A $.  Figure \ref{IOVRP} provides further insight by showing that $ k_{6m} $ is particularly constrained by matching to the observation of $ [CK] $ when POLARIS was overexpressed.


Figure \ref{IP2dsim} suggests that there is a positive trend between non-implausible values of $ k_{11}/k_{10} $, the ratio for the rate of ethylene receptor conversion from its active to inactive form, to the conversion rate from inactive to active form, and $ k_{13}/k_{12} $, the parameter representing the ratio of ethylene decay rate to biosynthesis rate.  This is consistent with current biological understanding that ethylene promotes the conversion from the active form of the ethylene receptor to its inactive form.  

The feeding terms $ \frac{V_{IAA}[IAA]}{Km_{IAA} + [IAA]} $, $ \frac{V_{CK}[cytokinin]}{Km_{CK} + [cytokinin]} $ and $ \frac{V_{ACC}[ACC]}{Km_{ACC} + [ACC]} $ are highly constrained by the measurements involving feeding, as can be seen by Figures \ref{VR1dsim} and \ref{IOVRP}.  In particular, the feeding of ethylene was constrained only after measurements involving the feeding of the ethylene hormone were measured.  Although this
is unsurprising, strong contradictions to such expected results may be an indication of a problem arising during the history matching procedure, hence these results are indicators that the history match was successful.  



In addition, our history matching procedure can also be used to investigate specific aspects of the model.  For example, the consequences of two experimentally determined, but opposing, regulatory relationships on constraining the non-implausible parameter space can be determined.  Our analysis, summarised in Figures \ref{OP1dsim}, \ref{IOVRP}, \ref{BoxplotsSim} and \ref{IP2dsimdisw14}, reveals what can be learnt about if further investigation was performed into the trend for $ f_c\_Auxin $.  In particular, we reveal the differences that a confirmed positive or negative trend for this output would have upon constraining the non-implausible parameter space.

In this article, we have developed the study of complex systems biology models using Bayes linear uncertainty analysis, with particular application to an important hormonal crosstalk model of Arabidopsis root development.  We have demonstrated the advantages of utilising a formal statistical model to link the biological model to reality.  We have also shown that performing a careful history match using implausibility measures, with the assistance of emulators, allows a global exploration of the input parameter space of the model.  In particular, by introducing experimental observations to the history matching procedure sequentially, we can explore constraints imposed by each group of observations, thus aiding the understanding of connections between the inputs and outputs of the model.  This in turn allows specific scientific learning objectives to be realised in the context of the model and by the links between the model and the biological system.  Being able to understand the contribution of particular experiments for informing us about acceptable input combinations can allow sensible experimental observations to be made relevant to the specific scientific learning objectives of the future.

Plant root developments are regulated by multiple hormones in a coordinated way.  Understanding the interdependence of the hormonal regulatory relationships, proteins and gene functions involved in root development is a difficult task.  We demonstrate that a combination of experimental observations, a model of hormonal crosstalk in Arabidopsis root development, and a Bayesian history match is able to establish the relationships between physical experiments and parameter space.  Thus, following the methodology we have developed in this work, future research should be able to more rationally integrate experimental measurements, model development and determination of non-implausible parameter space, for elucidating the complexity in hormonal signalling systems \citep{CCACE, HTDPWCM}.

\begin{acknowledgement}
 JL and KL gratefully acknowledge the Biotechnology \& Biological Sciences Research Council (BBSRC) for funding in support of this study. SEJ is in receipt of an EPSRC studentship. IV gratefully acknowledges MRC and EPSRC funding. 
\end{acknowledgement}

\appendix

\section*{Appendices}

\section{Extra Parameter $ \lambda $ \label{EPlambda} }

In order to perform a full analysis on the Arabidopsis model, we introduced a parameter $ \lambda = V_i/V_m $ to represent the ratio of cytosolic volume $ V_i $ to the volume of the cell wall $ V_m $.  This section outlines why it is necessary to introduce this parameter.

All but one of the outputs of the model represent chemical concentrations within the interior of the cell.  Therefore, relative values of chemical volume and chemical concentration of these outputs are the same.  On the other hand, however, $ PIN1pm $ should represent the concentration of PIN protein in the exterior of the cell.  The volume of the membrane is less than the volume of the interior of the cell, and this needs to be taken into account.  We outline an appropriate method to address this issue given the approximate nature of the model, which represents a single cell and membrane.  We need conservation of mass to hold for the overall mass of the PIN protein, or equivalently for flux into the membrane to be equal to flux out of the membrane.  This can be represented by:
\begin{equation}  
\frac{d[PIN1pm]}{dt} =  \lambda\left( k_{1_v24}[PIN1pi] - \frac{k_{25a}[PIN1pm]}{\displaystyle 1 + \frac{[Auxin]}{k_{25b}}} \right) \label{Equil}
\end{equation} 
where $ \lambda = V_i/V_m $ represents the ratio of the volume of the interior of an average cell $ V_i $ to the volume of the exterior of an average cell $ V_m $.  In general, we can introduce $ \lambda $ as an additional model parameter to the model equation for $ [PIN1pm] $ as given by \ref{Equil}, however since we are modelling at equilibrium only, we can instead just incorporate the effects of this extra model parameter into the rate parameters $ k_3 $ and $ k_{25a} $, leaving the equations unaltered.  Essentially we investigate $ k_3 = \frac{k_3^\prime}{\lambda} $ and $ k_{25a} = \frac{k_{25a}^\prime}{\lambda} $, where $ k_3^\prime $ and $ k_{25a}^\prime $ are the rate parameters assuming equal variance in both cell and membrane.

There are several ways to treat the additional parameter $ \lambda $.  $ \lambda $ could be varied as an extra parameter to the history match over a range of values believed to correspond to cell interior-membrane volume ratios.  The effect of varying cell sizes is already a feature of model discrepancy.  We therefore believed that it was adequate to fix $ \lambda $ and incorporate the uncertainty of $ \lambda $ as a source of internal model discrepancy \citep{MDEIMD, GFBUA}.  Internal model discrepancy refers to aspects of the model discrepancy which can be informed about by running the model.  Expert elicitation about the ratio $ \lambda $ led us to fix $ \lambda = 6 $ and suggested that a reasonable range of possible values for $ \lambda $ was $ [2,16] $.  We made sure that model discrepancy arising from varying the value of $ \lambda $ was captured in each output's model discrepancy assessment.

\printnomenclature

\bibliographystyle{authordate1-nospace}

\end{document}